\documentclass{vldb}
\usepackage{graphicx}
\usepackage{balance} 
\usepackage{enumitem}

\usepackage{tcolorbox}
\usepackage{courier}
\usepackage[scaled]{helvet}
\usepackage{url}
\usepackage{listings}
\usepackage{enumitem}
\usepackage{algorithm}
\usepackage{graphicx}
\usepackage{comment}
\usepackage{amsmath}
\usepackage{amssymb}
\usepackage{bbm, dsfont}
\usepackage{mathrsfs}
\usepackage{xcolor}
\usepackage{tikz}
\usepackage{pgfplots}
\usepackage{esvect}
\usepackage{xspace}
\usepackage{array, multirow, multicol}
\usepackage{rotating, makecell} % cell rotation
\usepackage[noend]{algpseudocode}
\usepackage{balance}
\usepackage{mathpartir}
\usepackage{stmaryrd}
\usepackage{mhchem}
\usepackage{wrapfig}
\usepackage{caption}
\usepackage{subcaption}
\usepackage{colortbl}
\usepackage{arydshln}
\usepackage{mathtools}

\newcommand{\mitra}{{\sc Mitra}\xspace}
\newcommand{\mitracore}{{\sc Mitra-core}\xspace}
\newcommand{\mitraxml}{{\sc Mitra-xml}\xspace}
\newcommand{\mitrajson}{{\sc Mitra-json}\xspace}

\ifdefined\comments
\newcommand{\navid}[1]{\textcolor{blue}{\textbf{NAVID:} #1}}
\newcommand{\isil}[1]{\textcolor{red}{\textbf{I\c{S}IL:} #1}}
\newcommand{\todo}[1]{\textcolor{red}{#1}}
\else
\newcommand{\navid}[1]{}
\newcommand{\isil}[1]{}
\newcommand{\todo}[1]{#1}
\fi

\newcommand{\tree}{\tau}
\newcommand{\pred}{\phi}
\newcommand{\textract}{\psi}
\newcommand{\cextract}{\pi}
\newcommand{\nextract}{\varphi}
\newcommand{\textracts}{\Psi}
\newcommand{\cextracts}{\Pi}
\newcommand{\nextracts}{\chi}
\newcommand{\logicop}{\unlhd}
\newcommand{\tuple}{t}

\newcommand{\lbb}{[\![}
\newcommand{\rbb}{]\!]}

\newcommand{\ex}{\mathcal{E}}
\newcommand{\rel}{\mathcal{R}}
\newcommand{\col}{\kappa}
\newcommand{\fta}{\mathcal{A}}
\newcommand{\preds}{\Phi}
\newcommand{\cost}{\theta}

\newcommand{\automaton}{\mathcal{A}}
\newcommand{\lang}{\mathcal{L}}
\newcommand{\states}{Q}

\newcommand{\finalstates}{Q_f}
\newcommand{\alphabet}{\mathcal{F}}
\newcommand{\transitionrules}{\Delta}
\newcommand{\rewriteto}{\rightarrow}

\newcommand{\semantics}[1]{\llbracket{#1}\rrbracket}

\newtheorem{example}{Example}
\newtheorem{definition}{Definition}
\newtheorem{theorem}{Theorem}

\newcommand{\irule}[2]%
   {\mkern-2mu\displaystyle\frac{#1}{\vphantom{,}#2}\mkern-2mu}

\algnewcommand{\IIf}[2]{\State\algorithmicif\ #1\ \algorithmicthen\ #2\ }
\algnewcommand{\IElseIf}[2]{\State\algorithmicelse\ \algorithmicif\ #1\ \algorithmicthen\ #2\ }
\algnewcommand{\IElse}[1]{\State\algorithmicelse\ #1\ }

\renewcommand{\dots}{\cdot\cdot}
\newcommand{\extree}{\texttt{T}}
\newcommand{\extuple}{\texttt{t}}
\newcommand{\exnode}{\texttt{n}}
\newcommand{\exbag}{\texttt{s}}
\newcommand{\bag}{s}

\makeatletter
\def\@copyrightspace{\relax}
\makeatother
\begin{document}

% ****************** TITLE ****************************************

\title{Automated Migration of Hierarchical Data to Relational Tables using Programming-by-Example}

% ****************** AUTHORS **************************************


\numberofauthors{3}

\author{
%% first
\alignauthor Navid Yaghmazadeh \\
\affaddr{University of Texas at Austin}\\
\email{nyaghma@cs.utexas.edu}
%% second
\alignauthor Xinyu Wang \\
\affaddr{University of Texas at Austin}\\
\email{xwang@cs.utexas.edu}
%% third
\alignauthor Isil Dillig \\
\affaddr{University of Texas at Austin}\\
\email{isil@cs.utexas.edu}
}

%\pagenumbering{roman}
%\input{cover-letter/cover-letter}
%\clearpage 

\maketitle

%\pagenumbering{arabic}

\begin{abstract}
While many applications export data in hierarchical formats like XML and JSON,  it is often necessary to convert such hierarchical documents to a relational representation. This paper presents a novel programming-by-example approach, and its implementation in a tool called \mitra, for automatically migrating tree-structured documents to relational tables. 
%Given a set of small input-output examples, our method automatically synthesizes a program that automates the desired task. 
We have evaluated the proposed technique using two sets of experiments. In the first experiment, we used \mitra to automate $98$ data  transformation tasks collected from StackOverflow. Our method can generate the desired program for $94\%$ of these benchmarks with an average synthesis time of $3.8$ seconds. In the second experiment, we used \mitra to generate programs that can convert real-world XML and JSON datasets to full-fledged relational databases. Our evaluation shows that \mitra can automate the desired transformation for all datasets.
\end{abstract}

\section{Introduction}\label{sec:intro}

Many applications store and exchange data using a hierarchical format, such as  XML or JSON documents. Such tree-structured data models are a natural fit  in cases where the underlying data is hierarchical in nature. Furthermore, since XML and JSON documents incorporate both data and meta-data, they are self-describing and portable. For these reasons, hierarchical data formats are  popular for exporting data and transferring them between different applications.

Despite the convenience of hierarchical data models, there are many situations that necessitate converting them to a relational format. %This transformation, sometimes referred to as ``shredding", may be necessary for a variety of reasons. 
For example, data stored in an XML document may need to be queried by an existing application that interacts with a relational database. Furthermore, because hierarchical data models are often not well-suited for efficient data extraction, %for  large datasets, 
converting them to a relational format is desirable when query performance is important.

\begin{comment}
Because of the widespread need to convert hierarchically structured data to a relational format, there has been significant research on XML and JSON shredding~\cite{DBforQXML-vldb99, xRel}. While many of these system use relational databases as an interface to store and query XML formats, they do not actually convert the underlying data to a relational representation. Hence, querying the database may still require traversing the underlying tree data structure. More general data exchange systems, such as Clio~\cite{clio,miller2000schema}, can convert  hierarchical data  to a relational format (e.g., based on user-specified mappings between elements in the source and target schema). However, because a schema mapping  may relate a source instance to many possible target instances, the resulting database instance may not be  what the user intended. 
\end{comment}

In this paper, we introduce a new technique based on \emph{programming-by-example (PBE)}~\cite{pbe1,pbe2} for converting hierarchically structured data to a relational format. In our methodology, the user provides a set of simple input-output examples to illustrate the desired transformation, and our system, \mitra~\footnote{stands for Migrating Information from Trees to RelAtions}, automatically synthesizes a program that performs the desired task. Because \mitra learns the target transformation from small input-output examples, it can achieve automation with  little guidance  from the user. In a typical usage scenario, a user would ``train" our system on a small, but representative subset of the input data  and then use the program generated by \mitra to convert a very large document to the desired relational representation.

While  programming-by-example has been an active research topic in recent years~\cite{flashfill,deepcoder,blinkfill,l2,hades,morpheus,foofah,qian2012sample}, most  techniques in this space focus on transformations between similarly structured data, such as string-to-string~\cite{flashfill,blinkfill}, tree-to-tree~\cite{hades,l2} or table-to-table transformations~\cite{morpheus,scythe,foofah,qian2012sample}. Unfortunately, automating transformations from tree- to table-structured data brings new technical challenges that are not addressed by prior techniques. First, because the source and target data representations are quite different, the required transformations are typically more complex than those between similarly-structured data.
Second, since each row in the target table corresponds to a relation between nodes in the input tree, the synthesizer needs to discover these ``hidden links" between  tree nodes.

%there may be relationships between arbitrary nodes of the tree, converting the tree to a relational format requires uncovering ``hidden links" between different nodes in the tree. 

\begin{comment}
\todo{
While  programming-by-example has been an active research topic in recent years~\cite{flashfill,deepcoder,blinkfill,l2,hades,morpheus,foofah,qian2012sample}, most  techniques in this space focus on transformations between similarly structured data, such as string-to-string~\cite{flashfill,blinkfill}, tree-to-tree~\cite{hades,l2} or table-to-table transformations~\cite{morpheus,scythe,foofah,qian2012sample}. Unfortunately, automating transformations from tree-structured to table-structured data brings new technical challenges that are not addressed by prior techniques. First, we need an expressive language; programs in this language should not only be able to produce table-structured data given tree-structured data as input, but more importantly, they should also be able to \emph{arrange} the data points from the input correctly as desired by the user (which requires uncovering ``hidden links" between different nodes in the input tree). Second, the language should be restricted enough and amenable to an efficient synthesis algorithm. 
}
\end{comment}

This paper addresses these challenges by presenting a new program synthesis algorithm that decomposes the synthesis task into two simpler subproblems that aim to learn the column and row construction logic seperately: 
\begin{itemize}[leftmargin=*]\itemsep0em
\item \emph{Learning the column extraction logic:} Given an attribute in a relational table, our approach first synthesizes a program to extract tree nodes that correspond to that attribute. In other words, we first ignore relationships between different tree nodes and construct each column separately. Taking the cross product of the extracted columns yields a table that  \todo{overapproximates} the target table \todo{(i.e., contains extra tuples)}.
\item \emph{Learning the row extraction logic:} Since the program learned in the first phase produces a table that overapproximates the target relation, the next phase of our algorithm synthesizes a program that filters out ``spurious" tuples generated in the first phase. In essence, the second phase of the algorithm discovers the ``hidden links" between different nodes in the original tree structure. 
\end{itemize}

\begin{comment}
\todo{
This paper addresses these challenges by presenting a novel domain-specific language that decomposes the transformation task into two subtasks (i.e., column and row construction logics) as well as an efficient synthesis algorithm that aims to solve these two subtasks separately: 
\begin{itemize}
\item \emph{Learning column construction logic:} 
Given an attribute in the output table, our approach first constructs a column corresponding to that given attribute by filling the column with data of nodes extracted from the input tree. In another word, our method first ignores relationships between different columns in the same row and constructs each column individually. Once all the columns are constructed, taking the cross product of them yields a table that always \todo{overapproximates} the target output table. 
\item \emph{Learning row construction logic:} Since the program learned in the first phase produces a table that \todo{overapproximates} tuples in the target  relation, the next phase of our algorithm synthesizes a program that filters out ``spurious" tuples generated by the first phase. In essence, the second phase of the algorithm discovers the ``hidden links" between different nodes in the original tree structure. 
\end{itemize}
}
\end{comment}

\begin{figure}
\begin{center}
\includegraphics[scale=0.25]{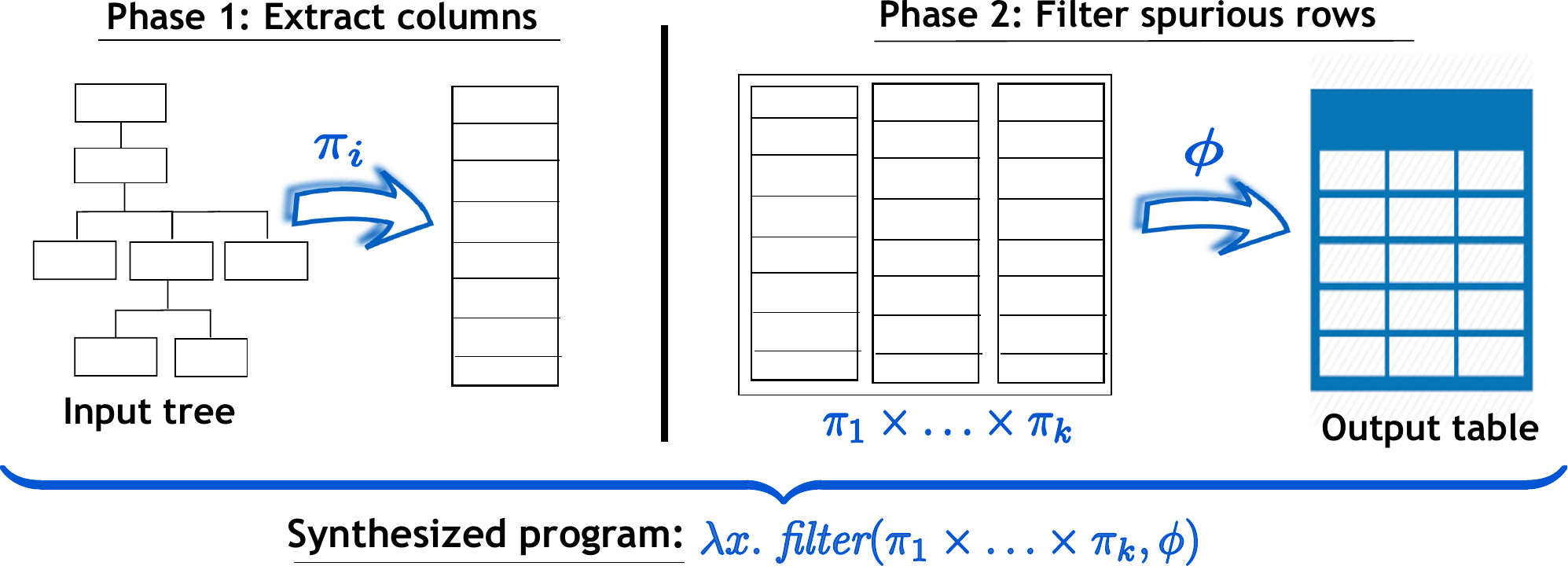}
\end{center}
\caption{Schematic illustration of our approach}\label{fig:overview}
\vspace{-0.15in}
\end{figure}

Figure~\ref{fig:overview} shows a schematic illustration of our synthesis algorithm. Given an input tree $T$ and output table $R$ with $k$ columns, our technique first learns $k$ different programs $\cextract_1, \ldots, \cextract_k$, where each \emph{column extraction} program $\cextract_i$ extracts from $T$ the data stored in the $i$'th column of $R$. Our synthesis algorithm then constructs an intermediate table by applying each $\cextract_i$ to the input tree and taking their cross product. Thus, the intermediate table $\cextract_1(T) \times \ldots \times \cextract_k(T)$ generated during the first phase  overapproximates the target table (i.e.,it may contain more tuples than $R$). In the next phase, our technique learns a \emph{predicate} $\pred$ that can be used to filter out exactly the spurious tuples from the intermediate table. Hence, the program synthesized by our algorithm is always of the form $\lambda x. \emph{filter}(\cextract_1 \times \ldots \times \cextract_k, \pred)$.  Furthermore, since the synthesized program should not be over-fitted to the user-provided examples, our method always learns the \emph{simplest} program of this shape that is consistent with the user-provided input-output examples. 

From a technical point of view, the contributions of this paper are three-fold. First, we propose a domain-specific language (DSL) that is convenient for expressing transformations between tree-structured and relational data.  Our DSL is expressive enough to capture many real-world data transformation tasks, and it also facilitates efficient synthesis by allowing us to decompose the problem into two simpler learning tasks. While the programs in this DSL may sometimes be inefficient, our method eliminates redundancies by memoizing shared computations in the final synthesized  program. This strategy allows us to achieve a good trade-off between expressiveness, ease of synthesis, and efficiency of the  generated programs.

The second technical contribution of this paper is a technique for automatically learning column extraction programs using \emph{deterministic finite automata (DFA)}. Given an input tree and a column from the output table, our method constructs a DFA whose language corresponds to the set of DSL programs that are consistent with this example. Hence, in our methodology, learning column extraction programs boils down  to finding a word (i.e., sequence of DSL operators) that is accepted by the  automaton. 
%In particular, such a tree corresponds to an abstract syntax tree (AST) of a DSL program that is consistent with the input-output examples. 

The third technical contribution of this paper is a novel technique for learning predicates that can be used to filter out spurious tuples in the intermediate table. Given a set of positive examples (i.e., tuples in the output table) and a set of negative examples (i.e., spurious tuples in the intermediate table), we need to find a smallest classifier in the DSL that can be used to distinguish the positive and negative examples. Our key observation is that this task can be reduced to  \emph{integer linear programming}. In particular, our method first  generates the universe of all possible atomic predicates over the DSL and then infers (using integer linear programming) \todo{a} \emph{smallest} subset of predicates that can be used to distinguish the positive and negative examples. Given such a subset, our method then uses standard logic minimization techniques to find a boolean combination of atomic predicates that can serve as a suitable classifier.

We have implemented our technique in a tool called \mitra, which consists of a language-agnostic synthesis core for tree-to-table transformations as well as  domain-specific plug-ins. While \mitra can generate code for migrating data from any tree-structured representation to a relational table, it requires plug-ins to translate the input format into our intermediate representation. Our current implementations contains two such plug-ins for XML and JSON documents. Furthermore, \mitra can be used to transform tree-structured documents to a full-fledged relational database by invoking it once for each table in the target database.

We have evaluated \mitra by performing two sets of experiments. In our first experiment, we use \mitra to automate 98 data transformation tasks collected from StackOverflow. \mitra can successfully synthesize the desired program for \todo{$94\%$} of these benchmarks, with an average synthesis time of \todo{$3.8$} seconds. In our second experiment, we use \mitra to migrate four real-world XML and JSON datasets (namely, IMDB, YELP, MONDIAL, and DBLP) to a full-fledged relational database. Our experiments show that \mitra can perform the desired task for all four datasets.

%XXX benchmarks for migrating tree-structured data to a relational format. Our experiments show that \mitra can learn the desired transformation for XXX\% of the benchmarks. Our experiments also show that learning programs using \mitra is practical; the average synthesis time in our experiments XXX seconds. Finally, we demonstrate that the programs synthesized by \mitra are efficient enough to migrate large XML and JSON documents to a relational format.

To summarize, this paper makes the following key contributions:

\vspace{-0.07in}
\begin{itemize}[noitemsep,leftmargin=*]
\item We propose a programming-by-example approach for migrating tree-structured documents to a relational format.
\item We propose a tree-to-table transformation DSL that faciliates synthesis by allowing us to decompose the synthesis task into two subproblems.
\item We describe a synthesis technique for learning column transformation programs using deterministic finite  automata.
\item We present a predicate learning algorithm that reduces the problem to a combination of integer linear programming and logic minimization.
\item We implement these ideas in a system called \mitra and empirically demonstrate its practicality and effectiveness.
\end{itemize}

\todo{The rest of the paper is organized as follows: We first provide an overview of our approach through a simple motivating example (Section~\ref{sec:overview}). We then introduce  \emph{hierarchical data trees} (Section~\ref{sec:prelim}) and our DSL for implementing tree-to-table transformations (Section~\ref{sec:language}). In Section~\ref{sec:synthesis}, we present our  synthesis algorithm and discuss our implementation in Section~\ref{sec:impl}. Finally, we present our empirical evaluation in Section~\ref{sec:eval} and survey related work in Section~\ref{sec:related}.}

\section{Overview}\label{sec:overview}

%\input{in-out-ex}

%% xml file example
%\begin{figure}
%\hspace{-0.15in}
%\includegraphics[scale=.25]{motiv-ex-xml}
%\caption{Input XML file}
%\label{fig:motiv-ex-xml}
%\end{figure}

\begin{figure}
\begin{subfigure}{.3\linewidth}
\centering
\includegraphics[scale=.2]{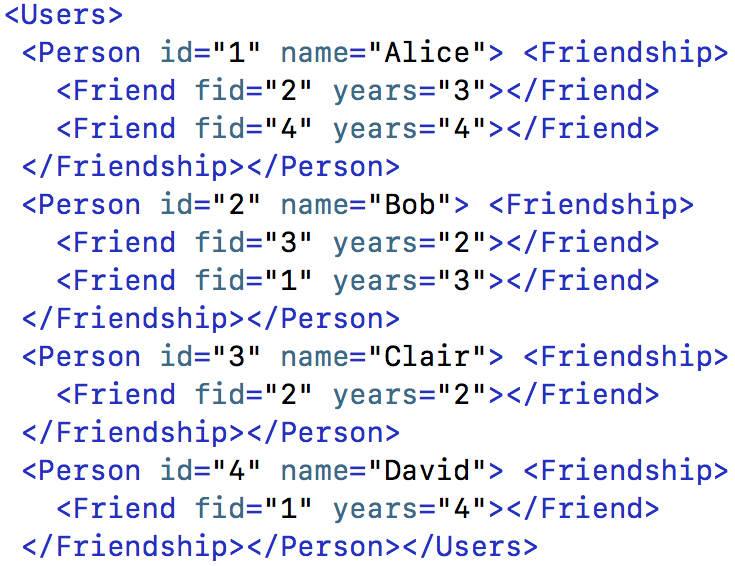}
\vspace{-0.22in}
\caption{Input XML}
\label{fig:motiv-ex-xml}
\end{subfigure}%
\hspace{1.1in}
\begin{subfigure}{.3\linewidth}
\centering
\vspace{0.2in}
\includegraphics[scale=.17]{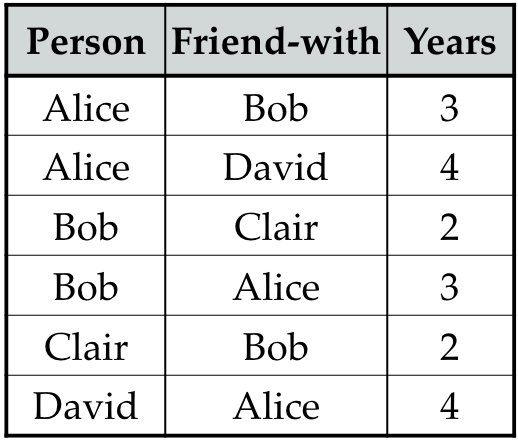}
\vspace{0.1in}
\caption{Output relation}
\label{fig:output-table}
\end{subfigure}
\vspace{5pt}
%\vspace{-0.02in}
\caption{Input-output example}
\label{fig:motiv-ex-in-out}
\vspace{-0.2in}
\end{figure}

In this section, we give a high-level overview of our technique with the aid of a simple motivating example. Consider the XML file from Figure~\ref{fig:motiv-ex-xml} which contains information about the users of a social network as well as the ``friendship" relation between them. Suppose a user wants to convert this XML document into the relational table shown in Figure~\ref{fig:output-table}. Observe that this transformation is non-trivial because the XML file stores this information as a mapping from each user to a list of friends, where each friend is represented by their {\tt fid}. In contrast, the desired table stores this information as tuples $(A, B, n)$, indicating that person with name \emph{A} is friends with user with name \emph{B} for $n$ years.

Suppose that the original XML file is much bigger than the one shown in Figure~\ref{fig:motiv-ex-xml}, so the user decides to automate this task by providing the input-output example from Figure~\ref{fig:motiv-ex-in-out} and uses \mitra to automatically synthesize the desired data migration program. 
We now give a brief overview of how \mitra generates the target program.

\begin{figure*}
\vspace{-5pt}
\begin{subfigure}{.7\linewidth}
\centering
\includegraphics[scale=.28]{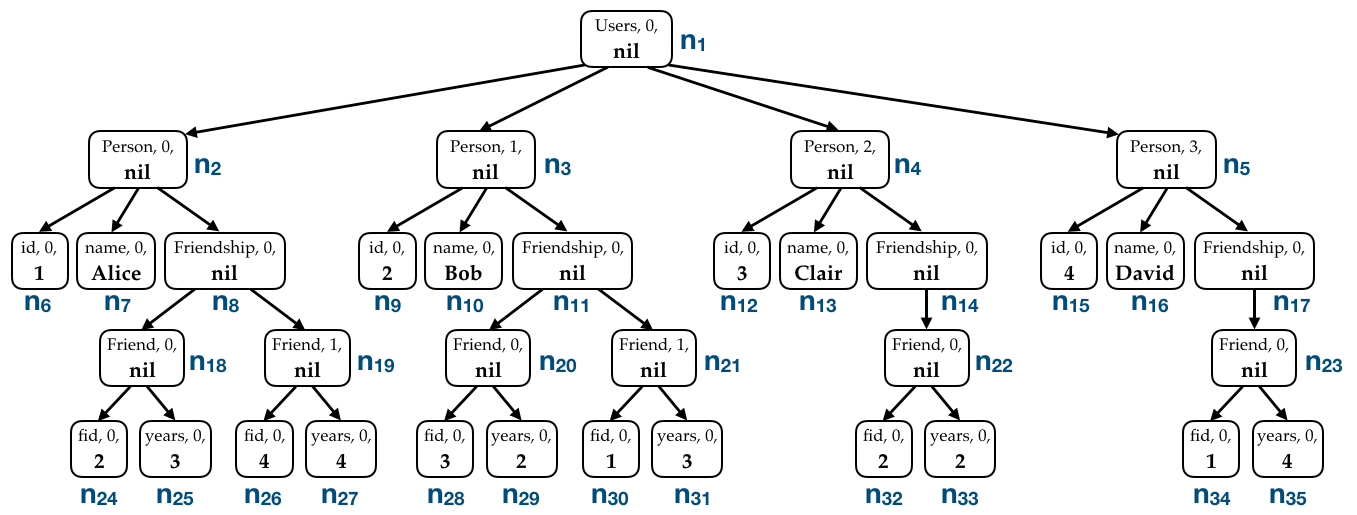}
\caption{\emph{Hierarchical data tree} representation of the input XML}
\label{fig:motiv-ex-hdt}
\end{subfigure}%
%\hspace{0.15in}
%\begin{subfigure}{.15\linewidth}
%\centering
%\vspace{0.45in}
%\includegraphics[scale=.18]{motiv-ex-rel}
%\vspace{0.1in}
%\caption{Desired Relation}
%\label{fig:motiv-ex-rel}
%\end{subfigure}%
\hspace{0.5in}
\begin{subfigure}{.18\linewidth}
\includegraphics[scale=.26]{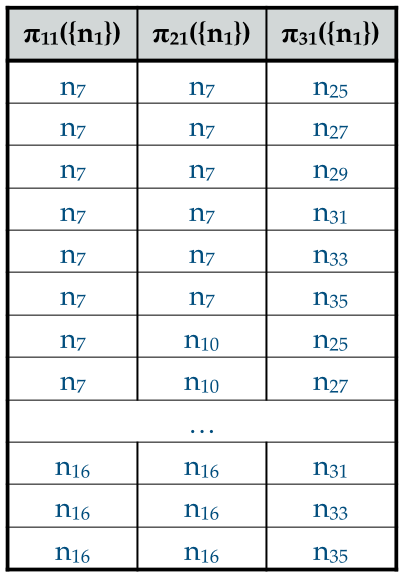}
\caption{Intermediate table}
\label{fig:motiv-ex-extracted}
\end{subfigure}
%\vspace{-0.02in}
\caption{Column extraction in the motivating example}
\label{fig:motiv-ex}
\vspace{-0.15in}
\end{figure*}

Internally, \mitra represents the input XML file as a \emph{hierarchical data tree}, shown in Figure~\ref{fig:motiv-ex-hdt}. Here, each node corresponds to an element in the XML document, and an edge from $n$ to $n'$ indicates that element $n'$ is inside $n$. The bold text  in each node from Figure~\ref{fig:motiv-ex-hdt} corresponds to the data stored in that element.

As mentioned in Section~\ref{sec:intro}, \mitra starts by learning all possible \emph{column extraction programs} that can be used to obtain column $i$ in the output table from the input tree. Figure~\ref{fig:motiv-ex-program} shows the extraction programs for each column. Specifically, \mitra learns a single column extraction program $\cextract_{11}$ (resp. $\cextract_{21}$) for the column {\tt Person} (resp. column {\tt Friend-with}). For instance, the program $\cextract_{11}$ first retrieves all children with tag {\tt Person} of the root node, and, for each of these children, it returns the child with tag {\tt name}. 
%Now let us look at how to extract the {\tt years} column in the output table. 
Since there are several ways to obtain the data in the {\tt years} column, \mitra learns  four different column extractors (namely, $\cextract_{31}, \ldots, \cextract_{34}$) for {\tt years}.

Next, \mitra \todo{conceptually} generates intermediate tables by applying each column extractor to the input tree and taking their cross product.~\footnote{Since this strategy may be inefficient, our implementation performs an optimization to eliminate the generation of intermediate tables in most cases. However, decomposing the problem in this manner greatly facilitates the synthesis task.} Since we have four different column extraction programs for the {\tt years} attribute, \mitra considers four different intermediate tables, one of which is shown in Figure~\ref{fig:motiv-ex-extracted}. In particular, this table is obtained using the \emph{table extraction} program $\textract$ presented in Figure~\ref{fig:motiv-ex-program}. Observe that  entries in the intermediate tables generated by \mitra refer to nodes from the input tree.

In the second phase of the synthesis algorithm, \mitra filters out spurious tuples in the intermediate table by learning a suitable \emph{predicate}. For instance, the intermediate table from Figure~\ref{fig:motiv-ex-extracted} contains several tuples that do not appear in the output table. As an example, consider the tuple $(n_7, n_7, n_{25})$. If we extract the data stored at these nodes, we would obtain the tuple {\tt (Alice, Alice, 3)}, which is not part of the desired output table. To filter out such spurious tuples from the intermediate table, \mitra learns a predicate $\pred$ such that $\pred$ evaluates to $\emph{true}$ for every tuple in the target table, and evaluates to $false$ for the spurious tuples. For our running example and the intermediate table from Figure~\ref{fig:motiv-ex-extracted}, \mitra learns the predicate $\pred_1 \land \pred_2$, where $\pred_1$ and $\pred_2$ are shown in  Figure~\ref{fig:motiv-ex-program}. Here, $\pred_1$ ensures that a tuple $(a, b, c)$ only exists if $a$ and $c$ are associated with the same person. Similarly, $\phi_2$ guarantees that $b$ refers to the person who has been friends with $a$ for $c$ years. For instance, since $\pred_2$ evaluates to false for the first tuple in the intermediate table from Figure~\ref{fig:motiv-ex-extracted}, this spurious tuple will be filtered out by the learnt predicate. 
%In the end, \mitra synthesizes the program $P$ shown in Figure~\ref{fig:motiv-ex-program}.

%the {\tt id} of the person from the {\tt friend-with} column is the same as the {\tt fid} of the person who has been a friend of $n_1$ for $n_3$ years.

%are children of a friend of the person whose name is used in column $1$, and $\pred_2$ guarantees that the $id$ of the person whose name is used in column $2$ is similar to the $fid$ of the friend which its child node is used in column $3$. 

%Clearly, $\textract_1$ contains tuples which do not appear in the target table (For instance, tuple $(n_7, n_7, n_{25})$ which corresponds to data tuple $(Alice, Alice, 3)$). Therefore, those ``spurious" tuples should be filtered out from the generated table. In order to do so, \mitra learns a predicate $\pred$ such that it evaluates to $true$ for every tuple in the target table, and evaluates to $false$ for the rest of tuples in $\textract_1$ which do not appear in the target table. The predicate $\pred = \pred_1 \land \pred_2$ learned by our algorithm is shown in the figure~\ref{fig:motiv-ex-prog}. Here, $\pred_1$ ensures that nodes in column $3$ are children of a friend of the person whose name is used in column $1$, and $\pred_2$ guarantees that the $id$ of the person whose name is used in column $2$ is similar to the $fid$ of the friend which its child node is used in column $3$. 

\begin{figure}
\vspace{-8pt}
\centering
\small
\[
\begin{array}{l}
\cextract_{11} : pchildren(children(\bag, Person), name, 0) \\
\cextract_{21} : pchildren(children(\bag, Person), name, 0) \\
\cextract_{31} : pchildren(children(\cextract_{F}, Friend), years, 0) \\
\cextract_{32} : pchildren(children(\cextract_{F}, Friend), fid, 0) \\
\cextract_{33} : pchildren(pchildren(\cextract_{F}, Friend, 0), years, 0) \\
\cextract_{34} : pchildren(children(\bag, Person), id, 0) \\
\cextract_{F} : pchildren(children(\bag, Person), Friendship, 0) \\[5pt]
\textract := (\lambda \bag. \cextract_{11}) \{ \emph{root}(\tree) \} \times (\lambda \bag. \cextract_{21}) \{ \emph{root}(\tree) \} \times (\lambda \bag. \cextract_{31}) \{ \emph{root}(\tree) \}\\ [5pt]
P \ := \lambda \tree. \ \emph{filter}(\textract, \lambda \tuple. \ \pred_{1} \land \pred_{2}) \\[5pt]
%\cextracts_1 := \{\cextract_{11} \}, \quad \cextracts_2 := \{\cextract_{21} \}, \quad \cextracts_3 := \{ \cextract_{31}, \cextract_{32}, \cextract_{33}, \cextract_{34} \} \\\ 
%\\
\begin{array}{ll}
\pred_{1} : &
((\lambda n. \ parent(n)) \ t[0]) \\
& = ((\lambda n. \ parent(parent(parent(n))) \ t[2]) \\
\end{array}
\\[7pt]
\begin{array}{ll}
\pred_{2} : & 
 ((\lambda n. \ child(parent(n), id, 0))) \ t[1]) \\ & = 
((\lambda n. \ child(parent(n), fid, 0)) \ t[2])\\
\end{array}
\end{array}
\]
\vspace{-5pt}
\caption{Synthesized program for the motivating example.}
\label{fig:motiv-ex-program}
\vspace{-0.15in}
\end{figure}

While all programs synthesized by \mitra are guaranteed to satisfy the input-output examples, not all of these programs may correspond to the user's intent. In particular, since examples are typically under-constrained, there may be multiple programs that satisfy the provided examples. For instance, in our running example, \mitra finds four different programs that are consistent with the examples. Our synthesis algorithm uses a heuristic ranking function based on \emph{Occam's razor} principle to rank-order the synthesized programs. For instance, programs that involve complex predicates are assigned a lower score by the ranking algorithm. Since the program $P$ shown in Figure~\ref{fig:motiv-ex-program} has the highest rank, 
%\mitra returns it as a solution. The user can now apply $P$ to the original (much larger) XML document and obtain the desired relational representation.
{ \mitra chooses it as the desired solution and optimizes $P$ by memoizing redundant computations and avoiding the generation of intermediate tables whenever possible. Finally, \mitra generates an executable XSLT program, which is available from \url{goo.gl/rcAHT4}.
%\todo{This optimization eliminates the generation of the intermediate overapproximated table in the synthesized program and highly increases its efficiency.} 
%\todo{Afterwards, the optimized program directly generates the output table, without going through an intermediate phase that produces an overapproximated table. Therefore, executing the optimized program on large real-world databases is highly efficient.}
The user can now apply the generated code to the original (much larger) XML document and obtain the desired relational representation. For instance, the synthesized program can be used \todo{to migrate} an XML document with more than 1 million elements to the desired relational table in $154$ seconds.} 

%Our algorithm iterates over all possible table overapproximations and generates a program $P_i$ for each of them by learning its corresponding filtering predicate. Finally, it ranks all those programs based the simplicity of column extractors and the filtering predicate, and returns the top ranked program as the desired transformation program. Program $P$ in figure~\ref{fig:motiv-ex-prog} shows the top ranked synthesized program by \mitra for this example.

\section{Preliminaries}\label{sec:prelim}

%In this section, we describe our representation of tree-structured documents and provide background on finite tree automata. 

%\subsection{Hierarchical Data Trees}

%% JSON file example
\begin{figure}[t]
%%\hspace{-0.15in}
\includegraphics[scale=.2]{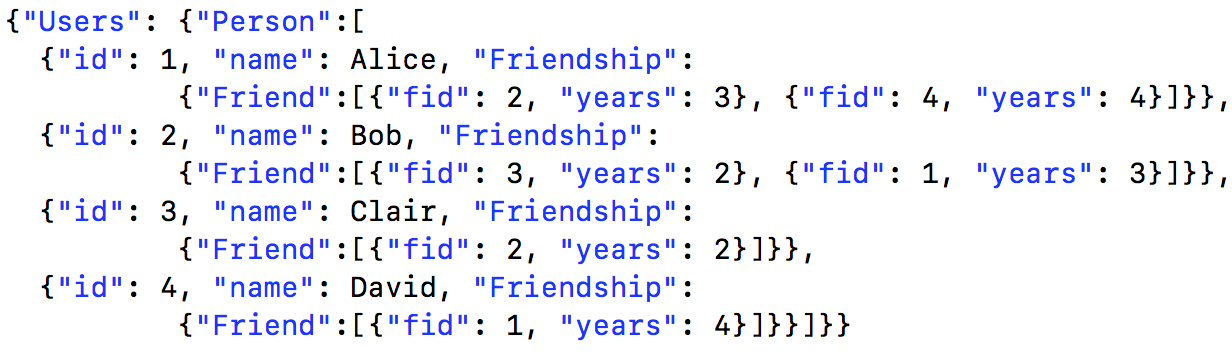}
\caption{Example of a JSON file}
\vspace{-0.1in}
\label{fig:json-ex}
\vspace{-0.1in}
\end{figure}

To allow for a uniform representation, we model tree-structured documents, such as XML, HTML and JSON, using so-called \emph{hierarchical data trees}, \todo{which is a variant of the data structure used in~\cite{hades}.}
\vspace{-0.05in}
\begin{definition} {\bf (Hierarchical Data Tree)} \label{def:hdt}
A hierarchical data tree (HDT) $\tree$ is a rooted tree represented as a tuple $\langle V, E \rangle$ where $V$ is a set of nodes, and $E$ is a set of directed edges. A node $v \in V$ is a triple $(tag, pos, data)$ where $tag$ is a label associated with  node $v$, $pos$ indicates that $v$ is the $pos$'th element with label $tag$ under $v$'s parent, and $data$ corresponds to the data stored at node $v$.
\end{definition}
\vspace{-0.05in}
Given a node $n = (t, i, d)$ in a hierarchical data tree, we use the notation $n.tag$, $n.pos$, and $n.data$ to indicate $t, i$, and $d$ respectively. We also use the notation $\emph{isLeaf}(n)$ to denote that $n$ is a leaf node of the HDT. In the rest of this paper, we assume that only leaf nodes contain data; hence, for any internal node $(t, i, d)$, we have $d=nil$. Finally, given an HDT $\tree$, we write $\tree.root$ to denote the root node of $\tree$.

 \vspace{0.1in}
\noindent
{\bf \emph{XML documents as HDTs}}.
We can represent  XML files as hierarchical data trees where nodes  correspond to XML elements. In particular, an edge from node $v'$ to $v = (t, i, d)$ indicates that the XML element $e$ represented by $v$ is nested directly inside  element $e'$ represented by $v'$. Furthermore since $v$ has tag $t$ and position $i$, this means $e$ is the $i$'th child with tag $t$ of $e'$. We also model XML attributes \todo{and text content} as nested elements. \todo{This design choice allows our model  to represent elements that contain a combination of nested elements, attributes, and text content.}
%and only leaf nodes of the HDT can store data. %Hence, for any internal node $v = (t, i ,d)$, we have $d = nil$.

%% xml file example
%\begin{figure}
%\hspace{-0.15in}
%\includegraphics[scale=.16]{xml-example}
%\caption{Example of an XML file}
%\label{fig:xml-ex}
%\end{figure}

\begin{comment}
%% JSON file example
\begin{figure}
\hspace{-0.15in}
\includegraphics[scale=.2]{json-example}
\caption{Example of a JSON file}
\label{fig:json-ex}
\vspace{-0.15in}
\end{figure}

%% corresponding HST for both xml and json examples
\begin{figure*}
%\hspace{-0.65in}
\begin{center}
\includegraphics[scale=.35]{hdt-example}
\end{center}
\caption{HDT representing the JSON file from Figure~\label{fig:json-ex}}
\label{fig:hdt-ex}
\end{figure*}
\end{comment}

\begin{example}
Figure~\ref{fig:motiv-ex-hdt} shows the HDT representation of the XML file from Figure~\ref{fig:motiv-ex-xml}. Observe how attributes are represented as nested elements in the HDT representation.
\end{example}

\noindent
{\bf \emph{JSON documents as HDTs}}. JSON documents store data as a set of nested key-value pairs.
We can model JSON files as HDTs in the following way: Each node \todo{$n =  (t, i, d)$} in the HDT corresponds to a key-value pair $e$ in the JSON file, where \todo{$t$} represents the key and \todo{$d$} represents the value. Since values in JSON files can be arrays, the position $i$ corresponds to the $i$'th entry in the array. For instance, if the JSON file maps key $k$ to the array $[18,45,32]$, the HDT contains three nodes of the form $(k, 0, 18)$, $(k, 1, 45), (k, 2, 32)$.
An edge from $n'$ to $n$ indicates that the key-value pair $e$ represented by $n$ is nested inside the key-value pair $e'$ represented by $n'$. 

\begin{example}\label{ex:json-to-tree}
Figure~\ref{fig:json-ex} shows the JSON document corresponding to the HDT representation in Figure~\ref{fig:motiv-ex-hdt}. Observe that JSON objects and arrays are represented as internal nodes with data \emph{nil}. For a given node $n$, we have $n.pos =0$ unless the parent of $n$ corresponds to an array. 
\end{example}

\section{Domain-Specific Language}\label{sec:language}

In this section, we present our domain-specific language (DSL) for implementing transformations from hierarchical data trees to relational tables. 
As standard, we represent relational tables as a bag of tuples, where each tuple is represented using a list. Given a relational table $\rel$, we use the notation $\emph{column}(\rel, i)$ to denote the $i$'th column of $\rel$, and we use the terms ``relation" and ``table" interchangably.

\begin{comment}

\begin{figure}
\centering
\[
\begin{array}{r c l}
{\rm Program} \ P & := & \lambda \tree. \ \emph{filter}(\textract, \lambda \tuple. \ \pred) %\ \emph{where} \ r = \{root(\tree) \} \\
\\ \ \\
{\rm Table \ Extractor} \  \textract & := & \cextract(\{\emph{root}(\tau)\}) \ | \ \textract_1 \times \textract_2 \\
%\ \emph{where} \ \ r = \{root(\tree) \} \\
%& & \textract_1 \times \textract_2 \\ 
\\ 
{\rm Column \ Extractor} \ \cextract & := & \lambda N. \ N \ 
\ |    \ \lambda N. \ children(\cextract(N), tag) \\
& & | \  \ \lambda N. \ pchildren(\cextract(N), tag, pos)   \\  \\
{\rm Node \ Extractor} \ \nextract & := &  \lambda n. \ n \ | \ \lambda n. \ parent(\nextract(n)) \\ 
& & | \ \ \lambda n. \ child(\nextract(n), tag, pos) \\  \\
{\rm Predicate} \ \pred & := & \nextract(nth(\tuple, c_1)) \logicop v \\
& & | \ \nextract_1(nth(t, c_1)) \logicop \nextract_2(nth(\tuple, c_2)) \\
&  & | \ \pred_1 \land \pred_2 \ | \ \pred_1 \lor \pred_2 \ | \ \neg \pred 
\end{array}
\]
%\vspace{-0.1in}
\caption{Syntax of  DSL for tree-to-table transformations. Here, $c$ is an integer constant, $v$ is a constant value (string, int, etc), and $\logicop$ is a logical operand such as $=, \leq$ etc.}
\label{fig:dsl-syntax}
\end{figure}

\end{comment}

\begin{figure}
\centering 
\small 
\[
\begin{array}{r c l}
{\rm Program} \ P & := & \lambda \tree. \emph{filter}(\textract, \lambda \tuple. \pred) %\ \emph{where} \ r = \{root(\tree) \} \\
\\[5pt]
{\rm Table \ Extractor} \  \textract & := & (\lambda \bag. \cextract) \ \{ \emph{root}(\tree) \} \ | \ \textract_1 \times \textract_2 \\[5pt]
{\rm Column \ Extractor} \ \cextract & := & \bag  \ | \ children(\cextract, tag) \\
& \ \ | & pchildren(\cextract, tag, pos)  \\ 
& \ \ | & descendants(\cextract, tag)  \\[5pt]
{\rm Predicate} \ \pred & := & \big( (\lambda n. \nextract) \ \tuple[i] \big) \logicop c \\
& \ \ | & \big( (\lambda n. \nextract_1) \ \tuple[i] \big) \logicop \big( (\lambda n. \nextract_2) \ \tuple[j] \big) \\ 
& \ \ | & \pred_1 \land \pred_2 \ | \ \pred_1 \lor \pred_2 \ | \ \neg \pred  \\ [5pt]
{\rm Node \ Extractor} \ \nextract & := &  n \ | \  parent(\nextract) \ | \ child(\nextract, tag, pos) 
\end{array}
\]
\caption{Syntax of our DSL. Here, $\tree$ denotes the input tree, $\tuple$ is bound to a tuple of nodes in $\tree$, and $n$ denotes a node in $\tree$. Furthermore, $i$ and $j$ are integers, and $c$ is a constant value (string, integer, etc). $\tuple[i]$ gives the $i$-th element in tuple $\tuple$.}
\label{fig:dsl-syntax}
\vspace{-0.2in}
\end{figure}

\begin{figure*}
\vspace{-5pt}
\centering
\scriptsize
\[
\begin{array}{r c l}
\lbb \ \emph{filter}(\textract, \lambda \tuple. \pred) \ \rbb_{\extree} & = & \{ \ (\exnode_1.\text{data}, \dots, \exnode_k.\text{data}) \ | \ \extuple \in \lbb \textract \rbb_{\extree}, \ \extuple =  (\exnode_1, \dots, \exnode_k), \ \lbb \pred \rbb_{\extuple, \extree} = \text{True} \ \} \\
\lbb \ \textract_1 \times \textract_2 \ \rbb_{\extree} & = & \{ \ (\exnode_1, \dots, \exnode_k, \exnode'_1, \dots, \exnode'_l) \ | \ (\exnode_1, \dots, \exnode_k) \in \lbb \textract_1 \rbb_{\extree}, \ (\exnode'_1, \dots, \exnode'_l) \in \lbb \textract_2 \rbb_{\extree} \ \} \\
\lbb \ (\lambda \bag. \cextract) \ \{ \text{root}(\tree) \} \ \rbb_{\extree} & = & \{ \ \exnode \ | \  \ \exnode \in \lbb \cextract \rbb_{\exbag, \texttt{T}}, \exbag = \{ \extree.\text{root} \} \ \}\\
\\
\lbb \ x \ \rbb_{\exbag, \texttt{T}} & = & \exbag \quad \quad \text{where } \exbag \text{ is a set of nodes in } \texttt{T} \\
\lbb \ children(\cextract, tag) \ \rbb_{\exbag, \texttt{T}} & = & \{ \ \exnode' \ | \ \exnode \in \lbb \cextract \rbb_{\exbag, \texttt{T}}, \ (\exnode, \exnode') \in \texttt{E}, \ \exnode'.\text{tag} = tag \ \} \ \text{where} \ \extree = (\texttt{V}, \texttt{E}) \\
\lbb \ pchildren(\cextract, tag, pos) \ \rbb_{\exbag, \texttt{T}} & = & \{ \ \exnode' \ | \ \exnode \in \lbb \cextract \rbb_{\exbag, \texttt{T}}, \ (\exnode, \exnode') \in \texttt{E}, \ \exnode'.\text{tag} = tag, \ \exnode'.\text{pos} = pos \ \} \ \text{where} \ \extree = (\texttt{V}, \texttt{E}) \\
\lbb \ descendants(\cextract, tag) \ \rbb_{\exbag, \texttt{T}} & = & \{ \ \exnode_{z} \ | \ \exnode_{1} \in \lbb \cextract \rbb_{\exbag, \texttt{T}}, \ \exists \{\exnode_{2},...,\exnode_{z-1}\} \subset \texttt{V} \ \text{s.t.} \ \forall {1 \leq x < z}. \  (\exnode_{x}, \exnode_{x+1}) \in \texttt{E}, \ \exnode_{z}.\text{tag} = tag \ \} \ \text{where} \ \extree = (\texttt{V}, \texttt{E}) \\
\lbb \  \big( (\lambda n. \nextract) \ \tuple[i] \big) \logicop c  \  \rbb_{\extuple, \extree} & = & \exnode^\prime.\text{data} \logicop c  \quad \quad \quad \quad \quad \  \text{where} \ \extuple = (\exnode_1, \dots, \exnode_l) \text{ and } \exnode^\prime = \lbb \nextract \rbb_{\exnode_i, \extree}  \\  \ \\ 
\lbb \  \big( (\lambda n. \nextract_1) \ \tuple[i]  \big) \logicop \big(  (\lambda n. \nextract_2) \ \tuple[j]  \big) \ \rbb_{\extuple, \extree} & = & \left\{ 
\begin{array}{ll}
\exnode_1^{\prime}.\text{data} \logicop \exnode_2^{\prime}.\text{data} & \text{if} \  \exnode_1^\prime \ \text{and} \ \exnode_2^{\prime} \text{ are both leaf nodes of } \extree \\ 
\exnode_1^{\prime} = \exnode_2^{\prime} & \text{if} \ \logicop \ \text{is } ``=" \text{, and neither } \exnode_1^{\prime} \text{ nor } \exnode_2^{\prime} \text{ is a leaf node of } \extree \\ 
\text{False} & \text{otherwise} 
\end{array}
\right. \\[5pt]
& & \text{where} \ \extuple = (\exnode_1, \dots, \exnode_l) \text{ and } \exnode_1^{\prime} = \lbb \nextract_1 \rbb_{\exnode_i, \extree} \text{ and }  \exnode_2^{\prime} = \lbb \nextract_2 \rbb_{\exnode_j, \extree} \\  \ \\
\lbb \ \pred_1 \land \pred_2 \ \rbb_{\extuple, \extree} & = & \lbb \pred_1 \rbb_{\extuple, \extree} \ \land \ \lbb \pred_2 \rbb_{\extuple, \extree} \\
\lbb \ \pred_1 \lor \pred_2 \ \rbb_{\extuple, \extree} & = & \lbb \pred_1 \rbb_{\extuple, \extree} \ \lor \ \lbb \pred_2 \rbb_{\extuple, \extree} \\
\lbb \ \neg \pred \ \rbb_{\extuple, \extree} & = & \neg \lbb \pred \rbb_{\extuple, \extree} \\
\lbb \ n \ \rbb_{\exnode, \extree} & = & \exnode  \\[5pt]
\lbb \ parent(\nextract) \ \rbb_{\exnode, \extree} & = & \left\{ 
\begin{array}{ll}
\exnode' &  \text{if} \ (\exnode', \lbb \nextract \rbb_{\exnode, \extree}) \in \texttt{E} \\
\bot &  \text{otherwise}
\end{array}
\right.  \quad \quad \quad \text{where}  \  \extree = (\texttt{V}, \texttt{E})\\[7pt]

\lbb \ child(\nextract, tag, pos) \ \rbb_{\exnode, \extree} & = & \left\{ 
\begin{array}{ll}
\exnode' & \text{if} \ ( \lbb \nextract \rbb_{\exnode, \extree}, \exnode') \in  \texttt{E} \text{ and } \exnode'.\text{tag} = tag \text{ and } \exnode'.\text{pos}  = pos \\
\bot &  \text{otherwise}
\end{array}
\right.  \quad \quad \quad \text{where}  \  \extree = (\texttt{V}, \texttt{E})\\
\end{array}
\]
\caption{Semantics of our DSL.}
\label{fig:dsl-semantics}
%\vspace{-0.1in}
\end{figure*}

Figure~\ref{fig:dsl-syntax} shows the syntax of our DSL, and Figure~\ref{fig:dsl-semantics} gives its semantics. Before we explain the constructs in this DSL,  an important point is that our language is designed to achieve a good trade-off between expressiveness and efficiency of synthesis. That is, while
our DSL can express a large class of tree-to-table transformations that arise in practice, it is designed to make automated synthesis practical.

The high-level structure of the DSL follows our synthesis methodology of decomposing the problem into two separate column and row extraction operations. In particular, a program $P$ is  of the form $\lambda \tree. \emph{filter}(\textract, \lambda t. \phi)$, where $\tree$ is the input HDT and $\psi$ is a \emph{table extractor} that extracts a relation $\mathcal{R}'$ from $\tree$. As mentioned in Section~\ref{sec:intro}, the extracted table $\mathcal{R}'$ overapproximates the target table $\mathcal{R}$. Therefore,  the  top-level \emph{filter} construct uses a predicate $\phi$ to filter out tuples in $\mathcal{R}'$ that do not appear in $\mathcal{R}$.

A \emph{table extractor} $\textract$ constructs a table by taking the cartesian product of a number of columns, where an entry in each column is a ``pointer" to a node in the input HDT.
Each column is obtained by applying a \emph{column extractor} $\cextract$ to the root node of the input tree. 
A column extractor $\cextract$ takes as input a set  of nodes and an HDT, and returns a set of HDT nodes. Column extractors are defined recursively and can be nested inside each other: The base case  simply returns the input set of nodes, and the recursive case  extracts other nodes from each input node using the \emph{children}, \emph{pchildren}, and \emph{descendants} constructs. Specifically, the \emph{children} construct extracts all children with a given tag, whereas \emph{pchildren} yields all children with a given tag and specified position. In contrast, the \emph{descendants} construct returns all descendants with the given tag. The formal (denotational) semantics of each  construct is given in Figure~\ref{fig:dsl-semantics}.

\begin{figure*}[t]
\begin{subfigure}{.25\linewidth}
\centering
\includegraphics[scale=.24]{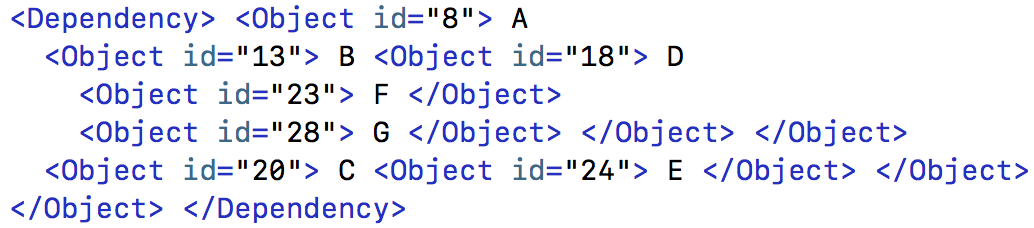}
\caption{Input XML}
\label{fig:ex2-xml}
\end{subfigure}%
\hspace{1.4in}
\begin{subfigure}{.2\linewidth}
\centering
\includegraphics[scale=.5]{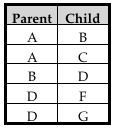}
\caption{Output relation}
\label{fig:ex2-rel}
\end{subfigure}%
\hspace{-0.3in}
\begin{subfigure}{.35\linewidth}
\centering
\vspace{-10pt}
\centering
\small
\[
\begin{array}{l}
P \ := \lambda \tree. \ \emph{filter}(\textract, \lambda \tuple. \ \pred_{1} \land \pred_{2}) \\[5pt]
\textract := (\lambda \bag. \cextract_{1}) \{ \emph{root}(\tree) \} \times (\lambda \bag. \cextract_{2}) \{ \emph{root}(\tree) \} \\[5pt]
\cextract_{1} = \cextract_{2} = pchildren(descendants(\bag, Object), text, 0) \\[5pt]
\pred_{1} : 
((\lambda n. \ child(parent(n), id, 0))) \ t[0]) < 20 \\
\pred_{2} : 
((\lambda n. \ parent(n)) \ t[0]) = ((\lambda n. \ parent(parent(n)) \ t[1]) \\
\\
\end{array}
\]
\vspace{-12pt}
\caption{Synthesized program}
\label{fig:ex2-program}
\end{subfigure}
\vspace{5pt}
\caption{Input-output example and the synthesized program for Example~\ref{ex:another}}
\label{fig:ex2-in-out}
\vspace{-0.15in}
\end{figure*}

Let us now turn our attention to predicates $\pred$ that can be used in the top-level \emph{filter} construct. Atomic predicates without boolean connectives are obtained by comparing the data stored in an HDT node with a constant $c$ or the data stored in another tree node. In particular, predicates make use of \emph{node extractors} $\nextract$ that take a tree node as input and return another tree node. Similar to column extractors, node extractors are recursively defined and can be nested within each other. Observe that node extractors allow accessing both parent and children nodes; hence, they can be used to extract any arbitrary node within the HDT from a given input node. (Figure~\ref{fig:dsl-semantics} gives the formal semantics).

Going back to the definition of predicates, $\pred$ takes a tuple $t$ and evaluates to a boolean value indicating whether $\tuple$ should be kept in the output table.  The simplest predicate $\big( (\lambda n. \nextract) \ \tuple[i] \big) \logicop c$ first extracts the $i$'th entry $\exnode$ of $\tuple$ and then uses the node extractor $\nextract$ to obtain another tree node $\exnode'$ from $\exnode$. This predicate evaluates to true iff $\exnode'.data \logicop c$ is true. The semantics of $\big( (\lambda n. \nextract_1) \ \tuple[i] \big) \logicop \big( (\lambda n. \nextract_2) \ \tuple[j] \big)$ is similar, except that it compares values stored at two tree nodes $\exnode, \exnode'$. In particular, if $\exnode, \exnode'$ are both leaf nodes, then we check whether the relation $\exnode.\text{data} \logicop \exnode'.\text{data}$ is satisfied. If they are internal nodes and the operator $\logicop$ is equality, then we check whether $\exnode, \exnode'$ refer to the same node. Otherwise, the predicate evaluates to false. 
More complex predicates are obtained using the standard boolean connectives $\land, \lor, \neg$.

\begin{comment}
\begin{example}
Figure~\ref{fig:motiv-ex-program} shows  the synthesized program for the input-output example given in Figure~\ref{fig:motiv-ex-in-out}. Column extractors use \emph{children} and \emph{pchildren} constructs to extract the desired nodes for each column. For instance, both $\cextract_{11}$ and $\cextract_{21}$ in Figure~\ref{fig:motiv-ex-program} first extract all children of the root of the tree with the tag $Person$, and then return their specific child node which has the tag $name$ and pos $0$. Predicate $\pred_1$ ensures that the parent of nodes in column $1$ and the third ancestor of node in column $3$ refer to the same node in the input tree. On the other hand, predicate $\pred_2$ checks the equality between the value of the $id$ sibling of nodes in column $2$ with the value of the $fid$ sibling of nodes in column $3$.
\end{example}

\end{comment}

%\input{ex2-hdt}

\begin{example}
Consider the data transformation task illustrated in Figure~\ref{fig:ex2-in-out}, in which we want to map the $text$ value of each $object$ element with $id$ less than $20$ in the XML file to the $text$ value of its immediate nested $object$ elements. %Figure~\ref{fig:ex2-hdt} represents the hierarchical data tree of the input XML file.
Figure~\ref{fig:ex2-program} shows the DSL program for this transformation. Here, column extractors $\cextract_1, \cextract_2$ use the \emph{descendants} and \emph{pchildren} constructs to extract all  children nodes with  tag $text$ and pos $0$ of any $object$ node reachable from the root. Predicate $\pred_1$ filters out all  tuples where the first element in the tuple has an \emph{id} sibling with value greater than or equal to $20$. The second predicate $\pred_2$ ensures that the  second element in the tuple is directly nested inside the first one.
%that the sibling node with tag $id$ and pos $0$ of the node in column $1$ has data value greater than or equal to $20$. $\pred_2$ ensures that the second ancestor of the node in column $2$ be the parent of the node in column $1$.
\label{ex:another}
\end{example}

\vspace{-0.1in}
\section{Synthesis Algorithm}\label{sec:synthesis}

In this section, we present our synthesis algorithm for converting an HDT into a relational table from input-output examples. Our technique can be used to transform an XML or JSON document into a relational database 
%by running our algorithm to synthesize a program that can produce each table in the target database. 
by running the algorithm once for each  table in the target database.

The top-level structure of our synthesis algorithm is given in Algorithm~\ref{alg:synth}. The algorithm takes a set of 
input-output examples of the form $\{ \extree_1 \rightarrow \rel_1, \dots, \extree_m \rightarrow \rel_m \}$, where each $\extree_i$ represents an HDT (input example) and $\rel_i$ represents its desired relational table representation (output example). Since the schema of the target table is typically fixed in practice, we assume that all  output tables (i.e., $\rel_i$'s) have the same number of columns. Given these input-output examples, the procedure {\sc LearnTransformation} returns a program $P^*$ in our DSL that is consistent with all  input-output examples. Furthermore, since $P^*$ is the simplest DSL program that satisfies the specification, it is expected to be a general program that is not over-fitted to the  examples.

\begin{comment}

Our synthesis algorithm requires  all output examples (i.e., $\rel_i$'s) to have the same number of columns. Please note that this assumption is realistic because the target schema is fixed. Given these input-output examples, the procedure {\sc LearnTransformation} returns a DSL program $P$ that is consistent with all  input-output examples. Furthermore, since $P$ is the simplest DSL program that satisfies the specification, it is expected to be a general program that is not over-fitted to the user-provided examples.

\end{comment}

As mentioned earlier, our methodology decomposes the synthesis task into two phases for extracting columns and rows. In the first phase, we synthesize a \emph{column extraction} program that yields an overapproximation of each column in the output table $\mathcal{R}$. The cartesian product of columns extracted by these column extraction programs further yields a table $\mathcal{R}'$ that overapproximates the target table $\mathcal{R}$. In the second phase, we learn a predicate $\pred$ that allows us to filter out exactly the ``spurious" tuples in $\mathcal{R}'$ that do not appear in the output table $\mathcal{R}$.

Let us now consider the {\sc LearnTransformation} procedure from Algorithm~\ref{alg:synth} in more detail. Given the examples, we first invoke a procedure called {\sc LearnColExtractors} that learns a \emph{set} $\cextracts_j$ of column extraction expressions such that applying any $\cextract \in \cextracts_j$ on each $\extree_i$ yields a column that overapproximates the $j$'th column in table $\mathcal{R}_i$ (lines 4-5). Observe that our algorithm learns a set of column extractors (instead of just one) because some of them might not lead to a desired \emph{filter} program. Our procedure for learning column extractors is based on deterministic finite automata and will be described in more detail in Section~\ref{sec:col-learn}.

\begin{comment}

For each of the $k$ columns in table $\mathcal{R}_i$, we first invoke a procedure called {\sc LearnColExtractors} that learns a \emph{set} $\cextracts_j$ of consistent column extraction expressions, i.e., applying any $\cextract \in \cextracts_j$ yields a column that overapproximates $\emph{column}(\mathcal{R}_i, j)$ (lines 6--7). Observe that our algorithm learns a \emph{set} of column extractors because there may be multiple such programs that are consistent with the input-output examples. Our procedure for learning column extractors uses finite tree automata and will be described in more detail in the next subsection.

\end{comment}

Once we have the individual column extraction programs for each column, we then obtain the set of all possible table extraction programs by taking the cartesian product of each column extraction program (line 6). Hence, applying each table extraction program $\textract \in \textracts$ to the input tree $\extree_i$  yields a table $\mathcal{R}'_i$ that overapproximates the output table $\mathcal{R}_i$.

\begin{comment}

\begin{algorithm}[t]
\caption{Synthesis Algorithm}
\label{alg:synth}
\begin{algorithmic}[1]

\vspace{0.1in}
\Procedure{LearnTransformation}{$\ex$}

\vspace{0.1in}
\State {\rm \bf Input:} Examples $\ex = \{ \tau_1 \rightarrow \rel_1, ..., \tau_m \rightarrow \rel_m \}$
\State {\rm \bf Output:} Synthesized program $P$
%\State {\rm \bf Requires:} All output tables in $\ex$ have $k$ columns
%\State {\rm \bf Ensures:} $\forall (\tree, \rel) \in \ex. \ P(\tree) = \rel$.

\vspace{0.1in}
\ForAll{$ 1 \le i \le k$ }
\State  $\Pi_i$ := {\sc LearnColExtractors}($\ex$, $i$); %\Comment{Learn all possible column extractors for each column in output table}
\EndFor
\vspace{0.05in}
\State $\Psi$ := $map(\Pi_1 \times ... \times \Pi_n, \lambda (\cextract_1, \cextract_2). \ \cextract_1 \times \cextract_2)$; 
\vspace{0.05in}
\ForAll {$\textract \in \Psi$ }
\State $\pred$ := {\sc LearnPredicate}($\ex$, $\textract$);
\If{$\pred \neq \bot$} 
\State \Return $\lambda \tree. \ \emph{filter}(\textract, \lambda t. \ \pred)$; 
\EndIf
\EndFor
\vspace{0.05in}
\State \Return $\bot$;
\EndProcedure
\end{algorithmic}
\end{algorithm}

\end{comment}

\begin{algorithm}[t]
\caption{Top-level synthesis algorithm}
\label{alg:synth}
\begin{algorithmic}[1]
\small
\vspace{4pt}
\Procedure{LearnTransformation}{$\ex$}

\vspace{4pt}
\State {\rm \bf Input:} Examples $\ex = \{ \extree_1 \rightarrow \rel_1, \dots, \extree_m \rightarrow \rel_m \}$. 
\State {\rm \bf Output:} Simplest DSL program $P^*$. 
%\State {\rm \bf Requires:} Each table $\rel_i$ has $k$ columns. 
%\State {\rm \bf Ensures:} $\forall i \in [1, m]: \lbb P^* \rbb_{\extree_i} = \rel_i$.

\vspace{4pt}
\ForAll{$ 1 \le j \le k$ }
\State  $\Pi_j$ := {\sc LearnColExtractors}($\ex$, $j$); %\Comment{Learn all possible column extractors for each column in output table}
\EndFor
\vspace{2pt}
\State $\Psi$ := $\Pi_1 \times \dots \times \Pi_n$; 
\vspace{2pt}
\State $P^* $ := $\bot$;
\ForAll {$\textract \in \Psi$ }
\State $\pred$ := {\sc LearnPredicate}($\ex$, $\textract$);
\If{$\pred \neq \bot$} 
\State $P := \lambda \tree. \emph{filter}(\textract, \lambda t. \pred$);
\If{$\cost(P)$ $<$ $\cost(P^*$)}  \  $P^* := P$
\EndIf
\EndIf
\EndFor
\vspace{2pt}
\State \Return $P^*$;
\EndProcedure
\end{algorithmic}
\end{algorithm}
\setlength{\textfloatsep}{6pt}

The next step of the synthesis algorithm (lines 7-12) learns the predicate used in the top-level \emph{filter} construct. For each table extraction program $\textract \in \textracts$, we try to learn a predicate $\pred$ such that $\lambda \tree. \emph{filter}(\textract, \lambda t. \phi)$ is consistent with the examples. Specifically, the procedure {\sc LearnPredicate} yields a predicate  that allows us to filter out spurious tuples in $\lbb \textract \rbb_{\extree_i}$. If there is no such predicate, {\sc LearnPredicate} returns $\bot$. Our predicate learning algorithm uses integer linear programming to infer a \emph{simplest} formula with the minimum number of atomic predicates. We describe the {\sc LearnPredicate} algorithm in more detail in Section~\ref{sec:pred-learn}. 

Since there may be multiple DSL programs that satisfy the provided examples, our method uses the Occam's razor principle to choose between different solutions. In particular, Algorithm~\ref{alg:synth} uses a heuristic cost function $\theta$ to determine which solution is simpler (line 12), so it is guaranteed to return a program that minimizes the value of $\theta$. Intuitively, $\theta$ assigns a higher cost to programs that use complex predicates or column extractors. We discuss the design of the heuristic cost function $\theta$ in Section~\ref{sec:impl}.

\vspace{-0.05in}
\subsection{Learning Column Extraction Programs}\label{sec:col-learn}

\begin{comment}
Our algorithm for learning column extraction programs is based on the synthesis technique using finite tree automata (FTA)~\cite{dace}. 
From a high-level, given an example, our algorithm constructs an FTA that symbolically represents a set of programs all of which are ``consistent" with the example. In particular, given an input tree $\extree$ and the $i$'th column in the output table $\rel$, our method constructs an FTA whose language consists of exactly the set $\Pi$ of column extraction programs such that applying any program in $\Pi$ on $\extree$ yields a column that \emph{overapproximates} the $i$'th column in $\rel$. In presence of multiple examples, we construct a separate FTA for each example and then take their intersection. The language of the resulting FTA includes column extractors that are consistent with \emph{all} examples. 
The main difference between our FTA method and previous FTA method~\cite{dace} is in how we label the final states when constructing the FTA, and we will explain this in more detail in our FTA construction algorithm. 
\end{comment}

Our technique for learning column extraction programs is based on deterministic finite automata (DFA). Given a tree (input example) and a single column (output example), our method constructs a DFA whose language accepts exactly the set of column extraction programs that are consistent with the input-output example. If we have multiple input-output examples, we construct a separate DFA for each example and then take their intersection. The language of the resulting DFA includes column extraction programs that are consistent with \emph{all} examples.

Let us now look at the {\sc LearnColExtractors} procedure shown in Algorithm~\ref{alg:extractor} in more detail. It takes the set of input-output examples $\ex$ as well as a column number $i$ for which we would like to learn the extractor. The algorithm returns a set of column extraction programs $\cextracts$ such that for every $\cextract \in \cextracts$ and every input-output example $(\extree, \rel)$, we have $\lbb \cextract \rbb_{\{ \extree.\text{root} \}, \extree} \supseteq \emph{column}(\rel, i)$.

\begin{algorithm}[t]
\caption{Algorithm for learning column extractors}
\label{alg:extractor}
\begin{algorithmic}[1]
\small

\vspace{4pt}
\Procedure{LearnColExtractors}{$\ex$, $i$}

\vspace{4pt}
\State {\rm \bf Input:} Examples $\ex$ and column number $i$. 
\State {\rm \bf Output:} Set $\cextracts$ of column extractors. 
%\State {\rm \bf Ensures:} $ \forall \cextract \in \cextracts. \ \forall (\extree, \rel) \in \ex.$ \\ \ \ \ \ \ \ \ \ \ \ \ \ \ \ \ \ \ \ \  $\lbb \cextract \rbb_{\{ \extree.\text{root} \}, \extree} \supseteq \emph{column}(\rel, i)$. 

\vspace{4pt}
\State $\ex' := \{ (\extree, \col) \ | \ (\extree, \rel) \in \ex \land \kappa = \emph{column}(\rel, i) \}$ 
\vspace{3pt}

\State $\fta$ := {\sc ConstructDFA}($e_0$) where $e_0 \in \ex'$
\vspace{3pt}
\ForAll{$e \in \ex' \backslash \{e_0\} $}
\State $\fta'$ := {\sc ConstructDFA}($e$) 
\State $\fta$ := {\sc Intersect}($\fta, \fta'$)
\EndFor
\vspace{2pt}
\State \Return {\sc Language}($\fta$)
\EndProcedure
\end{algorithmic}
\end{algorithm}
\setlength{\textfloatsep}{6pt}

\begin{comment}
The {\sc LearnColExtractors} procedure is presented in Algorithm~\ref{alg:extractor}. It takes the set of input-output examples $\ex$ as well as a column $i$ for which we would like to learn the extractor. Its return value is a set of column extraction programs $\cextracts$ such that, for every $\cextract \in \cextracts$ and every input-output example $(\tree, \rel)$, we have $\cextract(\tree) \supseteq \emph{column}(\rel, i)$.

\end{comment}

Algorithm~\ref{alg:extractor} starts by constructing a  set of examples $\ex'$ mapping each input tree to the $i$'th column of the output table (line 4). Then, for each example $e \in \ex'$, we construct a DFA that represents all column extraction programs consistent with $e$ (lines 5-8). The set of programs consistent with all examples $\ex'$ corresponds to the language of the intersection automaton $\automaton$ from line 9. \todo{In particular,  the {\sc Intersect} procedure used in line 8 is the standard intersection procedure for DFAs~\cite{automata_theory}. Concretely, the intersection of two DFAs $\fta_1$ and $\fta_2$ only accepts programs that are accepted by both $\fta_1$ and $\fta_2$ and is constructed by taking the cartesian product of $\fta_1$ and $\fta_2$ and defining the accepting states to be all states of the form $(q_1, q_2)$ where $q_1$ and $q_2$ are  accepting states of $\fta_1$ and $\fta_2$ respectively.} 

\begin{figure}[t]
\vspace{-0.15in}
\[ 
\begin{array}{cr}
\irule{
\exbag = \{ \texttt{T}.\text{root} \}
}{
q_0 = q_{\exbag} \in Q 
}
& {\rm (1)}
\\ \\ 
\irule{
\begin{array}{cc}
q_{\exbag} \in Q  \quad  \texttt{tag} \text{ is a tag in } \extree  \\ 
\lbb \emph{children}(\bag, \texttt{tag}) \rbb_{\exbag, \extree} = \exbag' 
\end{array}
}{
q_{\exbag'} \in Q, \ \delta(q_{\exbag},  \emph{children}_{\texttt{tag}}) = q_{\exbag'} 
} 
& {\rm (2)}
\\ \\ 
\irule{
\begin{array}{cc}
q_{\exbag} \in Q  \quad  \texttt{tag} \text{ is a tag in } \extree  \quad  \texttt{pos} \text{ is a pos in } \extree  \\ 
\lbb \emph{pchildren}(\bag, \texttt{tag}, \texttt{pos}) \rbb_{\exbag, \extree} = \exbag'
\end{array}
}{
q_{\exbag'} \in Q, \ \delta(q_{\exbag},  \emph{pchildren}_{\texttt{tag}, \texttt{pos}})=  q_{\exbag'}
} 
& {\rm (3)}
\\ \\ 
\irule{
\begin{array}{cc}
q_{\exbag} \in Q \quad   \texttt{tag} \text{ is a tag in } \extree  \\ 
\lbb \emph{descendants}(\bag, \texttt{tag}) \rbb_{\exbag, \extree} = \exbag'
\end{array}
}{
q_{\exbag'} \in Q, \ \delta(q_{\exbag}, \emph{descendants}_{\texttt{tag}}) = q_{\exbag'}  
} 
& {\rm (4)}
\\ \\ 
\irule{
\exbag \supseteq \emph{column}(\rel, i)
}{
q_{\exbag} \in F 
}
& {\rm (5)}
\end{array}
\]
\caption{FTA construction rules. $\extree$ is the input tree, $\rel$ is the output table, and $i$ is the column to be extracted.}
\label{fig:fta-rules}
%\vspace{-0.05in}
\end{figure}

The key part of the {\sc LearnColExtractors} procedure is the {\sc ConstructDFA} method, which constructs a deterministic finite  automaton $\mathcal{A} = (Q, \Sigma, \delta, q_0, F)$ from an input-output example using the rules shown in Figure~\ref{fig:fta-rules}. Here, the states $Q$ of the automaton correspond to sets of nodes in the input HDT. We use the notation $q_{\exbag}$ to denote the state representing the set of nodes $\exbag$.  The alphabet $\Sigma$ corresponds to the names of column extraction functions in the DSL. Specifally, we have: 
\[ 
\begin{array}{lll}
\Sigma & = & \{ \emph{children}_{\texttt{tag}}  \ | \  \texttt{tag} \text{ is a tag in } \extree \}  \\ 
&  \cup  & \{ \emph{pchildren}_{\texttt{tag}, \texttt{pos}}  \ | \  \texttt{tag} \ (\texttt{pos}) \text{ is a tag (pos) in } \extree \} \\ 
& \cup & \{ \emph{descendants}_{\texttt{tag}}  \ | \  \texttt{tag} \text{ is a tag in } \extree  \}
\end{array}
\]

In other words, each symbol in the alphabet corresponds to a DSL operator (instantiated with labels and positions from the input HDT). Transitions in the DFA are constructed using the semantics of DSL operators: Intuitively, given a DSL construct $f \in \{ \emph{children}, \emph{pchildren}, \emph{descendants} \}$ and a state $q_\exbag$, the DFA contains a transition $q_\exbag \xrightarrow{f} q'_\exbag$ if applying $f$ to $\exbag$ produces $\exbag'$. The initial state of the DFA is $\{ \extree.root \} $, and we have  $q_\exbag \in F$ iff  $\exbag$   overapproximates the $i$'th column in table $\rel$.

Let us now look at the construction rules shown in Figure~\ref{fig:fta-rules} in more detail. 
Rules (1)-(4) process the column extractor constructs in our DSL and construct states and/or transitions. 
The first rule adds $q_{ \{ \extree.\text{root} \} }$ as an initial state because the root node of the HDT is directly reachable. The second rule adds a state $q_{\exbag'}$ and a transition $\delta(q_\exbag, \emph{children}_{\texttt{tag}}) = q_{\exbag'}$ if  \emph{children}($\exbag$, $\texttt{tag}$) evaluates to $\exbag'$. Rules (3) and (4) are similar to rule (2) and process the remaining column extraction functions \emph{pchildren} and \emph{descendants}. For example, we have $\delta(q_\exbag, \emph{pchildren}_{\texttt{tag}, \texttt{pos}} ) = q_{\exbag'}$ if $\emph{pchildren}(\exbag, \texttt{tag}, \texttt{pos})$ evaluates to $\exbag'$.
The last rule in Figure~\ref{fig:fta-rules} identifies the final states. In particular, a state $q_\exbag$ is a final state if $\exbag$ is a superset of the target column (i.e., output example).

\begin{theorem}\label{thm:fta}
\todo{
Let $\fta$ be the DFA constructed by Algorithm~\ref{alg:extractor} for a set of input-output examples $\ex$ and a column number $i$. Then, $\fta$ accepts a column extraction program $\cextract$ in our DSL iff \sloppy$\forall (\extree, \rel) \in \ex. \ \lbb \cextract \rbb_{\{ \extree.\text{root} \}, \extree} \supseteq \emph{column}(\rel, i)$.~\footnote{\todo{Proofs of all theorems are available in Appendix~\ref{appendix:proofs}.}}
}
\end{theorem}

%Different from previous FTA method~\cite{dace} that labels states as final iff they \emph{exactly match} the output example, we mark a state as final if it \emph{overapproximates} the output example. This is because our synthesis methodology uses an overapproximate decomposition of the problem. 

%\noindent 
%{\bf \emph{Remark.}} 
%Limit length of programs in FTA construction to ensure termination of the construction procedure. 

\begin{example}
Suppose  the DFA constructed using rules from Figure~\ref{fig:fta-rules} accepts the word $ab$ where $a = \emph{descendants}_{object}$ and $b = \emph{pchildren}_{text, 0}$. This word corresponds to the following column extractor program:
\[
\cextract = \emph{pchildren}(\emph{descendants}(s, \emph{object}), \emph{text}, 0)
\]
If the input example is $\extree$ and the output example is $\emph{column}(\rel, i)$, then we have $((\lambda s. \cextract) \ \{\extree.root\}) \supseteq \emph{column}(\rel, i)$.
%In our setting, being consistent with the input-output example means that $((\lambda x. \cextract) \ {\extree.root})$ is a superset of the target column in the output table.
\end{example}

\subsection{Learning Predicates}\label{sec:pred-learn}

We now turn our attention to the predicate learning algorithm {\sc LearnPredicate} shown in Algorithm~\ref{alg:pred}. This procedure takes the input-output examples $\ex$ and a candidate table extractor $\textract$ and returns a predicate $\pred$ such that for every $(\extree, \rel) \in \ex$, the program $\emph{filter}(\textract, \lambda t. \pred)$ yields the desired output table $\rel$ on input $\extree$.

The algorithm starts by constructing a (finite) universe $\Phi$ of all possible atomic predicates that can be used in formula $\pred$ (line 4). These predicates are constructed for a set of input-output examples $\ex$ and a candidate table extractor $\textract$ using rules from Figure~\ref{fig:pred-rules}. While these rules are not very important for understanding the key ideas of our technique, let us consider rule (4) from Figure~\ref{fig:pred-rules} as an example. According to this rule, we generate an atomic predicate $\big( (\lambda n. \nextract) \ \tuple[i] \big) \logicop c$ if $i$ is a valid column number in the range $[1, k]$, $c$ is a constant in one of the input tables, and $\nextract$ is a ``valid" node extractor for column $i$. Rules (1)-(3) in Figure~\ref{fig:pred-rules} define what it means for a node extractor $\nextract$ to be valid for column $i$, denoted as $\nextract \in \nextracts_i$. In particular, we say $\nextract$ is a valid node extractor for column $i$ if applying $\nextract$  does not ``throw an exception" (i.e., yield $\bot$) for any of the entries in the $i$'th column of the generated  intermediate tables. \footnote{Recall that these intermediate tables are obtained by applying $\textract$ to the input HDTs in $\ex$.}

\begin{algorithm}[t]
\caption{Algorithm for learning predicates}
\label{alg:pred}
\begin{algorithmic}[1]
\small

\vspace{4pt}
\Procedure{LearnPredicate}{$\ex$,  $\textract$}

\vspace{4pt}
\State {\rm \bf Input:} Examples $\ex$, a candidate table extractor $\textract$. 
\State {\rm \bf Output:} Desired predicate $\pred$. 
%\State {\rm \bf Requires:} $\forall (\extree, \rel) \in \ex. \ \rel \subseteq \lbb \textract \rbb_{\extree}$. 
%\State {\rm \bf Ensures:} $\forall (\extree, \rel) \in \ex. \ \lbb \emph{filter}(\textract, \lambda t. \pred) \rbb_{\extree} = \rel$. 

\vspace{4pt}
\State $\preds$ := {\sc ConstructPredUniverse}($\ex$, $\textract$) 
\State $\ex^+ := \emptyset; \ \ex^- := \emptyset$
\vspace{2pt}
\ForAll{$(\extree, \rel) \in \ex$}
\ForAll{$t \in \lbb \textract \rbb_{\extree} $}
\If{$t \in \rel$} 
\State $\ex^+ := \ex^+ \cup \{t\}$
\Else \ $\ex^- := \ex^- \cup \{t\}$
\EndIf
\EndFor
\EndFor

\vspace{2pt}
\State $\preds^*$ := {\sc FindMinCover}$(\preds, \ex^+, \ex^-)$
%\State \Return {\sc FindFormula}$(\preds^*, \ex^+, \ex^-)$
\vspace{2pt}
\State {\rm Find} $\phi$ such that:
\vspace{2pt}
\State \quad \quad  (1) $\phi$ is boolean combination of preds in $\preds^*$
\State \quad \quad (2) $\phi(t) = \left \{ 
\begin{array}{ll}
1 & \emph{if} \ \ t \in \ex^+ \\
0 & \emph{if} \ \ t \in \ex^-
\end{array}
\right .$
%\State \quad \quad (3) $\phi$ uses the minimum \# of  connectives
\vspace{2pt}
\State \Return $\pred$

\EndProcedure
\end{algorithmic}
\vspace{-0.05in}
\end{algorithm}
\setlength{\textfloatsep}{6pt}

\begin{algorithm}[t]
\vspace{-0.15in}
\caption{Algorithm to find minimum predicate set}
\label{alg:cover}
\begin{algorithmic}[1]
\small

\vspace{14pt}
\Procedure{FindMinCover}{$\preds$,  $\ex^+$, $\ex^-$}

\vspace{4pt}
\State {\rm \bf Input:} Universe of predicates $\preds$
\State {\rm \bf Input:} Positive examples $\ex^+$, negative examples $\ex^-$
\State {\rm \bf Output:} Set of predicates $\preds^* $ where $\preds^* \subseteq \preds$ 
%\State {\rm \bf Ensures:} \[ \forall e_1 \in \ex^+. \forall e_2 \in \ex^-. \exists \pred \in \preds^*. \pred(e_1) \neq \pred(e_2) \]

\vspace{4pt}
\ForAll{$(\pred_k, e_i, e_j) \in \preds \times \ex^+ \times \ex^-$}
%\ForAll{$\pred_k \in \preds$}
\vspace{2pt}
\State $a_{ijk} =\left \{ 
\begin{array}{ll}
1 & \emph{if} \  \pred_k(e_i) \neq \pred_k(e_j) \\
0 & \emph{otherwise}
\end{array}
\right . $
%\EndFor
\EndFor
\vspace{4pt}
\State \emph{minimize} $\sum\limits_{k=1}^{|\preds|} x_k$
\vspace{2pt}
\State \emph{subject to:} 
\State \quad \quad \quad $\forall (e_i, e_j) \in \ex^+ \times \ex^-.  \ \sum\limits_{k=1}^{|\preds|} a_{ijk} \cdot x_k \geq 1$ 
\State  \quad \quad \quad $\land \ \forall k \in [1, |\preds|]. \ x_k \in \{ 0, 1 \} $

\vspace{4pt}
\State \Return $\{ \pred_i \ | \ \pred_i \in \preds \land x_i = 1 \}$

\EndProcedure
\end{algorithmic}
\vspace{-0.05in}
\end{algorithm}
\setlength{\textfloatsep}{6pt}

The next step of {\sc LearnPredicate} constructs a set of \emph{positive} and \emph{negative} examples to be used in {\sc LearnPredicate} (lines 5--10). In this context,  positive examples $\ex^+$ refer to  tuples that should be present in the desired output table, whereas negative  examples $\ex^-$ correspond to  tuples that should be filtered out. The goal of the predicate learner is to find a formula $\pred$ over atomic predicates in $\Phi$ such that $\pred$ evaluates to true for all positive examples and to false for all negative examples. In other words, formula $\pred$ serves as a classifier between $\ex^+$ and $\ex^-$.

To learn a suitable classifier, our algorithm first learns a \emph{minimum} set of atomic predicates that are necessary for distinguishing the $\ex^+$ samples from the $\ex^-$ ones. Since  our goal is to synthesize a general program that is not over-fitted to the input-output examples, it is important that the synthesized predicate $\pred$ is as simple as possible.
We formulate the problem of finding a simplest classifier as a combination of \emph{integer linear programming (ILP)} and \emph{logic minimization}~\cite{mccluskey,quine}. In particular, we use ILP to learn a minimum set of predicates $\Phi^*$ that must be used in the classifier, and then use circuit minimization techniques to find a DNF formula $\pred$ over $\Phi^*$ with the minimum number of boolean connectives.

Our method for finding the minimum set of atomic predicates is given in Algorithm~\ref{alg:cover}. The {\sc FindMinCover} procedure takes as input the universe $\Phi$ of all predicates as well as  positive and negative examples $\ex^+$, $\ex^-$. It returns a subset $\Phi^*$ of $\Phi$ such that, for every pair of examples $(e_1, e_2) \in \ex^+ \times \ex^-$, there exists an atomic predicate $p \in \Phi^*$ such that $p$ evaluates to different truth values for $e_1$ and $e_2$ (i.e., $p$ differentiates $e_1$ and $e_2$). 

We solve this optimization problem by reducing it to 0-1 ILP in the following way: First, we introduce an indicator variable $x_k$ such that $x_k = 1$ if $p_k \in \Phi$ is chosen to be in $\Phi^*$ and $x_k = 0$ otherwise. Then, for each predicate $p_k$ and every pair of examples $(e_i, e_j) \in \ex^+ \times \ex^-$, we introduce a (constant) variable $a_{ijk}$ such that we assign $a_{ijk}$ to 1 if predicate $p_k$ distinguishes $e_i$ and $e_j$ and to $0$ otherwise. Observe that the value of each $a_{ijk}$ is known, whereas  the assignment to each variable $x_k$ is to be determined.

To find an optimal assignment to variables $\vec{x}$, we set up the 0-1 ILP problem shown in lines 7--10 of Algorithm~\ref{alg:cover}. First, we require that for every pair of examples, there exists a predicate $p_k$ that distinguishes them. This requirement is captured using the constraint in line 9: Since $a_{ijk}$ is $1$ iff $p_k$ distinguishes $e_i$ and $e_j$, the constraint at line 9 is satisfied only when we assign at least one of the $x_k$'s differentiating $e_i$ and $e_j$ to 1. The objective function at line 7 minimizes the sum of the $x_k$'s, thereby forcing us to choose the minimum number of predicates that are sufficient to distinguish every pair of positive and negative examples.

\begin{figure}[t]
\[
\small  
\begin{array}{cr}
\irule{
}{
\cextract_i, \ex \vdash n \in \nextracts_i 
}
& {\rm (1)}
\\ \\ 
\irule{
\begin{array}{cc}
\cextract_i, \ex \vdash \nextract \in \nextracts_i   \\ 
\forall \extree \rightarrow \rel \in \ex. \ \forall \exnode \in \cextract_i(\extree). \ \lbb  \emph{parent}(\nextract) \rbb_{\exnode, \extree} \neq \bot
\end{array}
}{
\cextract_i, \ex \vdash \emph{parent}(\nextract) \in \nextracts_i
} 
& {\rm (2)}
\\ \\ 
\irule{
\begin{array}{cc}
 \cextract_i, \ex \vdash \nextract \in \nextracts_i \\ 
%\texttt{tag} \text{ is a tag in } \extree \quad \quad \texttt{pos} \text{ is a pos in } \extree \\
\forall \extree \rightarrow \rel \in \ex. \ \forall \exnode \in \cextract_i(\extree). \ \lbb  \emph{child}(\nextract, \texttt{tag}, \texttt{pos}) \rbb_{\exnode, \extree} \neq \bot \\
\end{array}
}{
\cextract_i, \ex \vdash \emph{child}(\nextract, \texttt{tag}, \texttt{pos}) \in \nextracts_i
} 
& {\rm (3)}
\\ \\ 
\irule{
\begin{array}{cc}
\textract = \cextract_1 \times \ldots \times \cextract_k, \quad i \in [1,k] \\
\cextract_i, \ex \vdash \nextract \in \nextracts_i \\
%\quad \quad \logicop \in \{<, \leq, =, \geq, >\} \\ 
\exists (\extree, \rel) \in \ex. \ c \in data(\extree)
%c \text{ is a data in } \extree \in \{\extree_1, ..., \extree_m\}
\end{array}
}{
\textract, \ex \vdash \big( (\lambda n. \nextract) \ \tuple[i] \big) \logicop c \in \preds
} 
& {\rm (4)}
\\ \\ 
\irule{
\begin{array}{cc}
\textract = \cextract_1 \times \ldots \times \cextract_k, \quad i \in [1,k], \quad j \in [1,k] \\
 \cextract_i, \ex \vdash \nextract_1 \in \nextracts_i \\
 \cextract_j, \ex \vdash \nextract_2 \in \nextracts_j \\ 
\end{array}
}{
\textract, \ex \vdash \big( (\lambda n. \nextract_1) \ \tuple[i] \big) \logicop \big( (\lambda n. \nextract_2) \ \tuple[j] \big) \in \preds
}
& {\rm (5)}
\end{array}
\]
\caption{Predicate universe construction rules. $\ex$ is the input-output examples  and $\textract = \cextract_1 \times ... \times \cextract_k$ is the candidate table extractor. Here $\nextracts_i$ indicates a set of node extractors that can be applied to the nodes extracted for column $i$.}
\label{fig:pred-rules}
%\vspace{-0.1in}
\end{figure}

Going back to Algorithm~\ref{alg:pred}, the return value $\Phi^*$ of {\sc FindMinCover} (line 11) corresponds to a minimum set of predicates that must be used in the classifier; however, we still need to find a \emph{boolean combination} of predicates in $\Phi^*$ that differentiates $\ex^+$ samples from the $\ex^-$ ones. Furthermore, we would like to find the \emph{smallest} such boolean combination for \todo{two} reasons: (1) large formulas might hinder the generality of the synthesized program as well as its readability, and (2) large formulas would incur more overhead when being evaluated at runtime. 

We cast the problem of finding a smallest classifier over predicates in $\Phi^*$ as a \emph{logic minimization} problem~\cite{mccluskey,quine}. In particular, given a set of predicates $\Phi^*$, our goal is to find a smallest DNF formula $\pred$ over predicates in $\Phi^*$ such that $\pred$ evaluates to \emph{true} for any positive example and to \emph{false} for any negative example. To solve this problem, we start by constructing a (partial) truth table, where the rows correspond to examples in $\ex^+ \cup \ex^-$, and the columns correspond to predicates in $\Phi^*$. The entry in the $i$'th row and $j$'th column of the truth table is \emph{true} if predicate $p_j \in \Phi^*$ evaluates to true for example $e_i$ and  false otherwise. The target boolean function $\phi$ should evaluate to true for any $e^+ \in \ex^+$ and false for any $e^- \in \ex^-$. Since we have a truth table describing the target boolean function, we can use standard techniques, such as the Quine-McCluskey method~\cite{mccluskey,quine}, to find a smallest DNF formula representing classifier $\phi$.

\begin{figure}
\centering
\small
\[
\begin{array}{c|c c c c c c c}
 & \boldsymbol{\pred_{1}} & \boldsymbol{\pred_{2}} & \boldsymbol{\pred_{3}} & \boldsymbol{\pred_{4}} & \boldsymbol{\pred_{5}} & \boldsymbol{\pred_{6}} & \boldsymbol{\pred_{7}} \\
\hline
\boldsymbol{e_1^+} & true & true & false & false & true & true & false \\
\boldsymbol{e_2^+} & false & true & true & true & true  & false & true \\
\boldsymbol{e_3^+} & false & true & true & true & false  & false & false \\
\boldsymbol{e_1^-} & false & false & true & true & false  & false & false \\
\boldsymbol{e_2^-} & false & true & true & true  & false  & false & true \\
\boldsymbol{e_3^-} & true & false & true & false & false  & false & true \\
\end{array}
\]
\caption{Initial truth table for the predicate universe $\preds$, positive examples $\ex^+$ and negative examples $\ex^-$.}
\label{fig:pred-leran-init}
\vspace{-0.13in}
\end{figure}

\begin{figure}[t]
\centering
\small
\[
\begin{array}{c|c c c c c c c c c}
 & \boldsymbol{\upsilon_{11}} & \boldsymbol{\upsilon_{12}} & \boldsymbol{\upsilon_{13}} & \boldsymbol{\upsilon_{21}} & \boldsymbol{\upsilon_{22}} & \boldsymbol{\upsilon_{23}} & \boldsymbol{\upsilon_{31}} & \boldsymbol{\upsilon_{32}} & \boldsymbol{\upsilon_{33}} \\
\hline
\boldsymbol{\pred_{1}} & 1 & 1 & 0 & 0 & 0 & 1 & 0 & 0 & 1 \\
\rowcolor{blue!20}
\boldsymbol{\pred_{2}} & 1 & 0 & 1 & 1 & 0 & 1 & 1 & 0 & 1 \\
\boldsymbol{\pred_{3}} & 1 & 1 & 1 & 0 & 0 & 0 & 0 & 0 & 0 \\
\boldsymbol{\pred_{4}} & 1 & 1 & 0 & 0 & 0 & 1 & 0 & 0 & 1 \\
\rowcolor{blue!20}
\boldsymbol{\pred_{5}} & 1 & 1 & 1 & 1 & 1 & 1 & 0 & 0 & 0 \\
\boldsymbol{\pred_{6}} & 1 & 1 & 1 & 0 & 0 & 0 & 0 & 0 & 0 \\
\rowcolor{blue!20}
\boldsymbol{\pred_{7}} & 0 & 1 & 1 & 1 & 0 & 0 & 0 & 1 & 1 \\
\end{array}
\]
\caption{Values of $a_{ijk}$ assigned in line $6$ of Algorithm~\ref{alg:cover}. Here $\upsilon_{ij}$ corresponds to $(e_i, e_j) \in \ex^+ \times \ex^-$.}
\label{fig:ilp-table}
%\vspace{-0.1in}
\end{figure}

\begin{figure}[t]
\vspace{-0.05in}
\centering
\small
\[
\begin{array}{c|c c c|c}
 & \boldsymbol{\pred_{2}} & \boldsymbol{\pred_{5}} & \boldsymbol{\pred_{7}} & \boldsymbol{\pred_{?}} \\
\hline
\boldsymbol{e_1^+} & true & true & false & true  \\
\boldsymbol{e_2^+} & true & true & true & true  \\
\boldsymbol{e_3^+} & true & false & false & true \\
\hdashline
\boldsymbol{e_1^-} & false & false & false & false  \\
\boldsymbol{e_2^-} & true & false & true & false  \\
\boldsymbol{e_3^-} & false & false & true & false  \\
\end{array}
\]
\caption{Truth table constructed in lines $14-16$ of Algorithm~\ref{alg:pred} for Example~\ref{ex:ilp}.}
\label{fig:truth-table}
%\vspace{-0.1in}
\end{figure}

\vspace{-0.05in}
\begin{theorem}\label{thm:predicate}
\todo{
Given  examples $\ex$ and table extractor $\textract$ such that $\forall (\extree, \rel) \in \ex. \ \rel \subseteq \lbb \textract \rbb_{\extree}$, Algorithm~\ref{alg:pred} returns a smallest DNF formula $\pred$ such that $\forall (\extree, \rel) \in \ex. \ \lbb \emph{filter}(\textract, \lambda t. \pred) \rbb_{\extree} = \rel$ if such a formula exists in our DSL.
}
\end{theorem}

\vspace{-0.15in}
\begin{example}\label{ex:ilp}
{Consider  predicate universe $\preds = \{ \pred_1, ..., \pred_7 \}$, a set of positive examples $\ex^+ = \{ e_1^+, e_2^+, e_3^+\}$, and a set of negative examples $\ex^- = \{ e_1^-, e_2^-, e_3^-\}$ with the truth table given in Figure~\ref{fig:pred-leran-init}. Here, the entry at row $e_i$ and column $\pred_j$ of Figure~\ref{fig:pred-leran-init} is \emph{true} iff tuple $e_i$ satisfies predicate $\pred_j$. The goal of the predicate learner is to find a formula $\pred_{?}$ with the minimum number of atomic predicates $\pred_i \in \preds$ such that it evaluates to true for all positive examples and  to false for all negative examples. In order to do so, the  {\sc FindMinCover} procedure first finds the minimum required subset of atomic predicates $\preds^*$ as described in Algorithm~\ref{alg:cover}. Figure~\ref{fig:ilp-table} shows the values of $a_{ijk}$ assigned in line 6 of Algorithm~\ref{alg:cover}. In particular, the entry at row $\pred_i$ and column $v_{jk}$ of Figure~\ref{fig:ilp-table} is \emph{true} iff $a_{ijk}$ is true. The highlighted rows in the Figure~\ref{fig:ilp-table} indicate the predicates which are selected to be in $\preds^*$ using integer linear programming. After finding $\preds^* = \{ \pred_2, \pred_5, \pred_7 \}$, lines 12--14 Algorithm~\ref{alg:pred} generate the (partial) truth table as shown in Figure~\ref{fig:truth-table}. The smallest DNF formula that is consistent with this truth table is $\pred_5 \lor (\pred_2 \land \neg \pred_7)$, so our algorithm returns this formula as a classifier.}
\end{example}

%\subsection{Synthesis Algorithm Guarantees}\label{sec:guarantees}

\todo{
The following two theorems state the soundness and completeness of our algorithm:
}

\vspace{-0.05in}
\begin{theorem}{\bf (Soundness)}\label{thm:soundness}
\todo{
Given examples $\ex = \{ \extree_1 \rightarrow \rel_1, \dots, \extree_m \rightarrow \rel_m \}$, suppose that {\sc LearnTransformation}($\ex$) yields $P^*$.  Then,  $\forall  (\extree_i, \rel_i) \in \ex$, we have $\lbb P^* \rbb_{\extree_i} = \rel_i$.
}
\end{theorem}

\vspace{-0.12in}
\begin{theorem}{\bf (Completeness)}\label{thm:completeness}
\todo{
Suppose there is a DSL program consistent with  examples $\ex = \{ \extree_1 \rightarrow \rel_1, \dots, \extree_m \rightarrow \rel_m \}$. Then,  {\sc LearnTransformation}($\ex$) eventually returns a program $P^*$ such that $\forall i \in [1, m]: \ \lbb P^* \rbb_{\extree_i} = \rel_i$.
}
\end{theorem}

\todo{Finally, the following theorem states that our synthesis algorithm returns a simplest DSL program with respect to the cost metric $\theta$:}

\vspace{-0.05in}
\begin{theorem}{\bf (Simplicity)}\label{thm:simplicity}
\todo{
%Algorithm~\ref{alg:synth} returns a simplest DSL program (w.r.t. the cost function $\theta$ in Section~\ref{sec:impl}) that satisfies Theorem~\ref{thm:soundness}. In other words,  
Given examples $\ex$, Algorithm~\ref{alg:synth} returns a  DSL program $P^*$ such that for any program $P$ satisfying $\ex$, we have $\theta(P^*) \leq \theta(P)$.
%$\forall P \ s.t. \ (\forall i \in [1, m]: \ \lbb P \rbb_{\extree_i} = \rel_i) \ \land \ (\theta(P) < \theta(P^*))$.
}
\end{theorem}

\noindent
\todo{
{\bf \emph{Complexity.}}  Our algorithm has worst-case exponential time complexity with respect to the size of input-output examples, as \textit{integer linear programming} and \textit{logic minimization} for a given truth table are both NP-hard problems. However, \emph{in practice}, the complexity of our algorithm does  not come close to the worst-case scenario. A more detailed explanation of the empirical complexity of the proposed algorithm can be found in Appendix~\ref{appendix:complexity}.
}
%\todo{
%However, \emph{in practice}, the complexity of our algorithm is close to $O(m^2 \cdot k^2 \cdot n^2 \cdot r \cdot (\log n)^k)$, where $m$ is the number of input-output examples, $k$ and $r$ are (respectively) the number of columns and maximum number of rows in output tables, and $n$ is the maximum number of nodes in an input tree. More precisely, the {\sc LearnColExtractor} function takes $O(m \cdot n^2)$ time to generate column extractors for each column, and a small number of table extractors are generated in line $8$ of Algorithm~\ref{alg:synth}. The bottleneck of our system is the {\sc LearnPredicate} procedure, which generates $O(k^2 \cdot n^2)$ atomic predicates, $O(m \cdot r)$ positive examples, and $O(m \cdot (\log n)^k)$ negative examples. This results in an empirical complexity of $O(m^2 \cdot k^2 \cdot n^2 \cdot r \cdot (\log n)^k)$ for Algorithm~\ref{alg:pred}. % which determines the overall complexity of our synthesis algorithm.
%}

\section{Implementation}\label{sec:impl}

\begin{figure}
%\vspace{-0.1in}
\begin{center}
\includegraphics[scale=0.28]{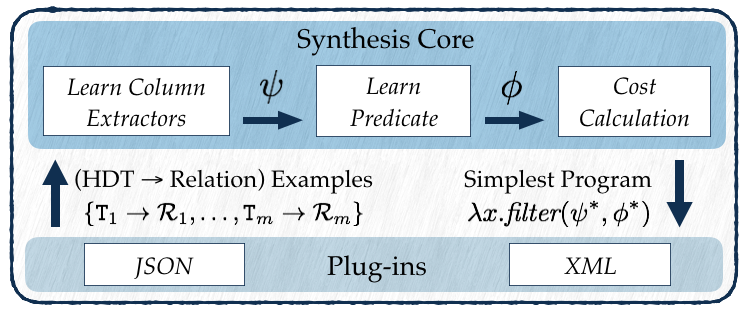}
\end{center}
\vspace{-0.1in}
\caption{Architecture of \mitra}\label{fig:layout}
%\vspace{-0.2in}
\end{figure}

We have implemented our synthesis algorithm in a tool called \mitra, which consists of approximately $7,000$ lines of Java code. \todo{As shown in Figure~\ref{fig:layout},} \mitra includes a domain-agnostic synthesis core (referred to as \mitracore) and a set of domain-specific plug-ins. Specifically, \mitracore accepts input-output examples in the form of (HDT, table) pairs and outputs a  program over the DSL shown in Figure~\ref{fig:dsl-syntax}. The goal of a \mitra plug-in is to  (1) convert the input document to our internal HDT representation, and (2) translate the  program synthesized by \mitracore to a target DSL. We have currently implemented two domain-specific plug-ins, called \mitraxml and \mitrajson, for XML and JSON documents respectively. Specifically, \mitraxml outputs programs written in XSLT, and \mitrajson generates JavaScript programs. \mitra can be easily extended to handle other forms of hierarchical documents \todo{(e.g., HTML and HDF)} by implementing suitable plug-ins.

\vspace{0.1in}\noindent
{\bf \emph{Cost function.}} Recall from Section~\ref{sec:synthesis} that our synthesis algorithm uses a heuristic cost function $\theta$ to choose the simplest program among all possible solutions that are consistent with the provided input-output examples. The cost function that we use in our implementation ranks programs based on the complexity of their predicates and column extractors and returns a program with the lowest cost. Given two programs $P_1, P_2$, our cost function assigns $P_1$ (resp. $P_2$) a lower cost if it uses fewer atomic predicates than $P_2$ (resp. $P_1$). If $P_1$ and $P_2$ use the same number of atomic predicates, then the cost function assigns a lower cost to the program that uses fewer constructs in the column extractors. 

\vspace{0.1in}\noindent
{\bf \emph{Program optimization.}} 
%\todo{ While our DSL is designed to facilitate synthesis,  programs in this language can be inefficient: In particular, the synthesized programs generate a (potentially large) intermediate table and then filter out the undesired tuples. To avoid inefficiencies caused by this design choice, \mitracore optimizes the synthesized programs by avoiding the generation of intermediate tables whenever possible. In particular, recall that our synthesized programs are of the form $\lambda x. \emph{filter}(\pi_1 \times \ldots, \pi_k, \phi)$, and suppose that all $\pi_i$'s  share the subprogram $\pi^*$. Our implementation exploits this shared program $\pi^*$ to avoid redundant computations. Furthermore, under certain conditions (determined by the relationship between predicate $\phi$ and shared subprogram $\pi^*$), we \centering an avoid the generation of intermediate tables altogether. A more detailed explanation of this optimization can be found in XXX.}
\todo{
While our DSL is designed to facilitate synthesis, programs in this language can be inefficient: In particular, the synthesized programs generate a (potentially large) intermediate table and then filter out the undesired tuples. To avoid inefficiencies caused by this design choice, \mitracore optimizes the synthesized programs by avoiding the generation of intermediate tables whenever possible. In particular, consider a synthesized program of the form $\lambda x. \emph{filter}(\pi_1 \times \pi_2, \phi)$. Instead of first taking the cross-product of $\pi_1, \pi_2$ and then filtering out undesired tuples, we optimize the synthesized program in such a way that the optimized program directly generates the final table by using $\phi$ to guide  table generation. More specifically, we use $\phi$ to find a prefix subprogram $\pi^*$ that is shared by $\pi_1, \pi_2$ with the property that any subsequent execution of the remaining parts of $\pi_1, \pi_2$ from any node in $\semantics{\pi^*}$ yields tuples that are guaranteed to satisfy $\phi$. Therefore, the optimized program avoids a post-filtering step and directly generates the output table. A more detailed explanation of this optimization can be found in Appendix~\ref{appendix:optimization}. 
}

%To generate a more efficient implementation, \mitracore identifies a maximally shared subprogram $\pi^*$ of each $\pi_i$ and emits the following code:
%\[
%\begin{array}{l}
%T := \emptyset \\
%\texttt{foreach} \ n \ \in \pi^*(\tau.\emph{root}) \\
%\ \ \ \ T := T \cup (\pi_1(n) \times \ldots \times \pi_k(n))\\
%\texttt{return} \ \emph{filter}(T, \phi) 
%\end{array}
%\]
%Essentially, this optimization allows us to reuse shared computations between each $\pi_i$, while also avoiding the generation of many spurious tuples that are subsequently eliminated using the \emph{filter} operation. }

%apply a given subprogram to the same tree node multiple times. To avoid inefficiencies arising from such redundant computation, the implementation of \mitracore memoizes intermediate computations. In particular, if two entries from the same row are obtained from a common ancestor in the input tree, the synthesized code re-uses the computation for extracting the common ancestor. 
%\todo{In essence, this optimization allows us to avoid generating an intermediate table most of the time. That is, rather than generating an intermediate table $T$ that overapproximates the target table and then filtering spurious tuples in $T$, this post-processing step often allows us to generate a program that fuses these two steps.}
%Since this optimization is performed at the IR level,  domain-specific plug-ins do not need to perform this kind of optimization to improve performance. 

\begin{table*}[!ht]
\small
\centering
\setlength\extrarowheight{1pt}
\newcolumntype{A}{ >{\centering\arraybackslash} m{0.07\textwidth} }
\newcolumntype{B}{ >{\centering\arraybackslash} m{0.055\textwidth} }
\newcolumntype{D}{ >{\centering\arraybackslash} m{0.14\textwidth} }
\begin{tabular}{|c|A|B|A|A|B|A|B|A|B|A|B|}
\hline
 & \multicolumn{3}{c|}{{\bf Benchmarks}} & \multicolumn{2}{c|}{{\bf Synthesis Time}} & \multicolumn{4}{c|}{{\bf Input-output Examples}} & \multicolumn{2}{c|}{{\bf Synth. Program}} \\
 \cline{2-12}
 & \multirow{2}{*}{{\bf \#Cols}} & \multirow{2}{*}{{\bf Total}} & \multirow{2}{*}{{\bf \#Solved}} & {\bf Median} & {\bf Avg.} & \multicolumn{2}{c|}{{\bf \#Elements}} & \multicolumn{2}{c|}{{\bf \#Rows}} & {\bf \#Preds} & {\bf LOC}\\
 \cline{7-10}
 & & & & {\bf (s)} & {\bf (s)} & {\bf Median} & {\bf Avg.} & {\bf Median} & {\bf Avg.} & {\bf (Avg.)} & {\bf (Avg.)} \\
\hline \hline
\multirow{5}{*}{\begin{turn}{90}\makecell{{\bf XML}}\end{turn}}
 & $\le$ 2  & 17 & 15 & 0.34 & 0.38 & 12.0 & 15.9 & 3.0 & 4.3 & 1.0 & 13.2 \\
\cline{2-12}
 & 3 & 12 & 12 & 0.63 & 3.67 & 19.5 & 47.7 & 4.0 & 3.8 & 2.0 & 17.2 \\
\cline{2-12}
 & 4 & 12 & 11 & 1.25 & 3.56 & 16.0 & 20.5 & 2.0 & 2.7  & 3.1 & 19.5 \\
\cline{2-12}
 & $\ge$ 5 & 10 & 10 & 3.48 & 6.80 & 24.0 & 27.2 & 2.5 & 2.6 & 4.1 & 23.3 \\
\cline{2-12}
 & {\bf Total} & {\bf 51} & {\bf 48} & {\bf 0.82} & {\bf 3.27}  & {\bf 16.5} & {\bf 27.2} & {\bf 3.0} & {\bf 3.5} & {\bf 2.4} & {\bf 17.8} \\
\cline{2-12}
\hline \hline

\multirow{5}{*}{\begin{turn}{90}\makecell{{\bf JSON}}\end{turn}}
 & $\le$ 2  & 11 & 11 &  0.12 & 0.27 & 6.0 & 7.4 & 2.0 & 2.7  & 0.9 & 21.3\\
\cline{2-12}
 & 3 & 11 & 11 & 0.48 & 1.13 & 7.0 & 10.5 & 3.0 & 3.5  & 2.0 & 23.0 \\
\cline{2-12}
 & 4 & 11 & 11 & 0.26 & 12.10 & 6.0 & 7.9 & 2.0 & 2.8  & 3.0 & 26.5 \\
\cline{2-12}
 & $\ge$ 5 & 14 & 11 & 3.20 & 3.85 & 6.0 & 8.1 & 2.0 & 2.5  & 4.9 & 28.0 \\
\cline{2-12}
 & {\bf Total} & {\bf 47} & {\bf 44}  & {\bf 0.31} & {\bf 4.33} & {\bf 6.0} & {\bf 8.5}  & {\bf 2.0} & {\bf 2.9} & {\bf 2.7} & {\bf 24.7} \\
\cline{2-12}
\hline \hline
\multicolumn{2}{|c|}{{\bf Overall}} & {\bf 98} & {\bf 92}  & {\bf 0.52} & {\bf 3.78}  & {\bf 11.0} & {\bf 18.7} & {\bf 3.0} & {\bf 3.2}& {\bf 2.6} & {\bf 21.6} \\
\hline
\end{tabular}
\vspace{0.1in}
\caption{Summary of our experimental evaluation}
\label{tab:result}
\vspace{-0.15in}
\end{table*}

\vspace{0.1in}\noindent
{\bf \emph{Handling full-fledged databases.}}  The synthesis algorithm that we described in Section~\ref{sec:synthesis} generates programs for converting a single HDT to a relational table. However, in practice, we would like to use \mitra to convert XML and JSON datasets to a complete database with a given schema. This transformation can be performed by invoking \mitra multiple times (once for each target table) and annotating primary and foreign keys of database tables.

\mitra ensures that the synthesized program obeys primary and foreign key constraints by performing a post-processing step.~\footnote{If the primary and foreign keys come from the input data set, we assume that the dataset already obeys these constraints. Hence, the following discussion assumes that the primary and foreign keys do not appear in the input dataset.} To ensure that the primary key uniquely identifies a given row in the table, the synthesized program generates primary keys as follows: If a given row in the database table is constructed from nodes $n_1, \ldots, n_k$ in the input tree, then we generate its primary key using an injective function $f(n_1, \ldots, n_k)$. Since each row in the table is constructed from a unique list of tree nodes, the generated primary key is guaranteed to be unique for each row as long as $f$ is injective.  In our implementation, $f$ simply concatenates  the unique identifiers for each tree node.

In order to ensure that a foreign key in table $T$ refers to a primary key in the table $T'$, the synthesized program for table $T$ must use the same function $f$ to generate the foreign keys. In particular, to generate the foreign key for a given row $r$ constructed from list of tree nodes $n_1, \ldots, n_k$, we need to find the tree nodes $n_1', \ldots, n_m'$ that are used to construct the corresponding row $r'$ in $T'$. For this purpose, we learn $m$ different (node extractor, node) pairs $(\nextracts_j, n_{t_j})$  such that $\nextracts_j(n_{t_j})$ yields $n_j'$ for all rows in the output examples for $T$ and $T'$. Finally, for a given row $r$ in $T$, we then generate the foreign key for $r$ as $f(\nextracts_1(n_{t_1}), \ldots,  \nextracts_m(n_{t_m}))$. This strategy ensures that the foreign and primary key constraints are satisfied as long as the learnt node extractors are correct.

\section{Evaluation}\label{sec:eval}

To evaluate \mitra, we perform experiments that are designed to answer the following questions:
\vspace{-3pt}
\begin{itemize}[noitemsep]
    \item[{\bf Q1.}] How effective is \mitra at synthesizing tree-to-table transformation programs? 
    \item[{\bf Q2.}] Can \mitra be used to migrate real-world XML and JSON datasets to the desired relational database?
    \item[{\bf Q3.}] Are the programs synthesized by \mitra fast enough to automate real-world data transformation tasks?
\end{itemize}

To answer these questions, we perform two sets of experiments: The first experiment evaluates \mitra on tree-to-table transformation tasks collected from StackOverflow, whereas the second experiment evaluates \mitra on real-world datasets. Both experiments are conducted on a MacBook Pro with 2.6 GHz Intel Core i5 processor and 8 GB of 1600 MHz DDR3 memory running OS X version 10.12.5.

\subsection{Accuracy and Running Time}\label{sec:eval-accuracy}

\noindent
\todo{
{\bf \emph{Setup.}}}
To perform our first experiment, we collected $98$ tree-to-table transformation tasks from StackOverflow by searching for relevant keywords (e.g., ``JSON to database", ``XML shredding", etc.).~\footnote{\todo{All  benchmarks are available from \cite{mitra_benchmarks}}}. Among these $98$ benchmarks, $51$ involve XML documents, while $47$ involve JSON files. 

Since \mitra requires input-output examples, we obtained the input XML/JSON file directly from the StackOverflow post. For output examples, we used the table provided in the StackOverflow post if one was present; otherwise, we constructed the desired output table based on the English description included in the post. 

%For each benchmark, we used \mitra to synthesize a program within a $60$ second time limit. 
%\todo{We chose a one minute time limit for the synthesis tasks because we believe a practical synthesis system should be able to generate the desired program within a reasonable period of time.}
\todo{For each benchmark, we used \mitra to synthesize a program that performs the given task.}
We manually inspected the tool's output to check whether the synthesized program performs the desired functionality. Even though any program synthesized by \mitra is guaranteed to satisfy the provided  input-output examples, it may not necessarily be the program that the user intended. Whenever the program synthesized by \mitra did not conform to the English description provided in the StackOverflow post, we updated the input-output examples to make them more representative. In our evaluation, we needed to update the original input-output example at most once, and the original examples were sufficient to learn the correct program {for the majority of  the benchmarks.}
%in $54.5\%$ of the benchmarks. \todo{Try to get this number to be higher!!}

\begin{table*}[t]
\small
%\scriptsize 
\centering
\setlength\extrarowheight{1pt}
\newcolumntype{A}{ >{\centering\arraybackslash} m{0.075\textwidth} }
\newcolumntype{B}{ >{\centering\arraybackslash} m{0.085\textwidth} }
\newcolumntype{D}{ >{\centering\arraybackslash} m{0.14\textwidth} }
\begin{tabular}{|c|A|A|B|B|A|A|A|A|A|}
\hline
\multirow{2}{*}{{\bf Name}} & \multicolumn{2}{c|}{{\bf Dataset}} & \multicolumn{2}{c|}{{\bf Database}} & \multicolumn{2}{c|}{{\bf Synthesis}} & \multicolumn{3}{c|}{{\bf Execution}} \\
 \cline{2-10}
 & {\bf Format} & {\bf Size} & {\bf \#Tables} & {\bf \#Cols} & {\bf Tot. Time(s)} & {\bf Avg. Time(s)}  & {\bf \#Rows} & {\bf Tot. Time(s)} & {\bf Avg. Time(s)} \\
 \hline \hline
{\bf DBLP} & XML & 1.97 GB& 9 & 39 & 7.41 & 0.82 & 8.312 M & 1166.44 & 129.60 \\
\hline
{\bf IMDB} & JSON & 6.22 GB & 9 & 35 & 33.53 & 3.72 & 51.019 M & 1332.93 & 148.10 \\
\hline
{\bf MONDIAL} & XML & 3.64 MB & 25 & 120 & 62.19 & 2.48 & 27.158 K & 71.84 & 2.87 \\
\hline
{\bf YELP} & JSON & 4.63 GB & 7 & 34 & 14.39 & 2.05 & 10.455 M & 220.28 & 31.46 \\
\hline
\end{tabular}
\vspace{0.1in}
\caption{Summary of using \mitra for migrating real-world datasets to a full DB. The columns labeled ``Tot. Time" include time for all database tables, whereas ``Avg. Time" is the average time per table.}
\label{tab:result-fullDB}
\vspace{-0.18in}
\end{table*}

\vspace{4pt}\noindent
\todo{
{\bf \emph{Results.}}
}
Table~\ref{tab:result} summarizes the results of evaluating \mitra on these $98$ benchmarks. The first part of the table provides information about our benchmarks, which we categorize into four classes depending on the number of columns in the target table.  Specifically, ``\#Cols" shows the number of columns in each category, and ``Total" shows the number of benchmarks in each category. The column labeled ``\#Solved" shows the number of benchmarks that \mitra was able to solve correctly. 
%within the $60$ second time limit. 
Overall, \mitra was able to synthesize the target program for \todo{$93.9\%$} of these benchmarks.

\begin{comment}
To understand \mitra's limitations, we investigated the $7$ benchmarks for which \mitra was not able to synthesize the desired program. We found that the desired program was not expressible in our DSL for $5$ of these $7$ benchmarks. For example, some of these $5$ benchmarks require the use of a conditional in the column extractor, which our DSL does not currently support. For the remaining $2$ benchmarks, there exists a DSL program that can perform the desired functionality, but \mitra was not able to find it within the $60$ second time limit. 
\todo{One of these $2$ benchmarks timed-out because of the abnormally large number of columns in the desired table ($26$ columns), and the other one timed-out because of the complexity of the desired program (which results in generating many possible table extractors and a relatively slow process of finding the desired predicate for each them).}
\end{comment}

%Going back to Table~\ref{tab:result}, 
The columns under ``Synthesis Time" show the median and average time that \mitra takes to synthesize the desired program in seconds. On average, \mitra synthesizes the target transformation in \todo{$3.8$} seconds, and the median time is even lower (\todo{$0.5$} seconds). We believe these results demonstrate that \mitra is quite practical.

Next, the section labeled  ``Input-output Examples" in Table~\ref{tab:result} describes properties of the provided input-output examples. The two columns under ``\#Elements" represent the median and average number of elements in the input document, and the``\#Rows" columns show the median and average number rows in the output table. Here, ``\#Elements" corresponds to the number of JSON objects and XML elements. As we can see from Table~\ref{tab:result}, the median number of elements in the input document is $11$, and the median number of rows in the input table is $3$. These results suggest that \mitra can synthesize the desired program from relatively small input-output examples.

The final section of Table~\ref{tab:result} describes properties of the synthesized programs. For instance, according to the``Preds" column,  the average number of atomic predicates used in predicates $\phi$ from our DSL is \todo{$2.6$}. More interestingly, the column labeled ``LOC" gives the number of lines of code in the synthesized XSLT and Javascript programs. On average, the size of the programs synthesized by \mitra is \todo{$21.6$}  (without including built-in functions, such as the implementation of \emph{getDescendants} or code for parsing the input file).

\vspace{4pt}\noindent
\todo{
{\bf \emph{Limitations.}}
}
To understand \mitra's limitations, we investigated the \todo{$6$} benchmarks for which \mitra was not able to synthesize the desired program. We found that the desired program was not expressible in our DSL for $5$ of these \todo{$6$} benchmarks. For example, some of these $5$ benchmarks require the use of a conditional in the column extractor, which our DSL does not currently support. 
\todo{For the other benchmark, there exists a DSL program that can perform the desired functionality, but \mitra 
ran out of memory due to the large number of columns in the target table.} 
%For the remaining $2$ benchmarks, there exists a DSL program that can perform the desired functionality, but \mitra was not able to find it within the $60$ second time limit. 
%\todo{One of these $2$ benchmarks timed-out because of the abnormally large number of columns in the desired table ($26$ columns), and the other one timed-out because of the complexity of the desired program (which results in generating many possible table extractors and a relatively slow process of finding the desired predicate for each them).}

\vspace{4pt}\noindent
\todo{
{\bf \emph{Performance.}}
}
We also evaluated the performance of the synthesized programs by running them on XML documents of size $512 \pm 20$ MB. In particular, we used the Oxygen XML editor~\cite{oxygen_xml} to generate XML files with a given schema and a specified number of elements. Among the $48$ XSLT programs generated using \mitra, $46$ of them were able to perform the desired transformation \todo{within approximately one minute.}.
% within a time limit of $300$ seconds.  
The running times of these $46$ programs range between $8.6$ and $65.5$ seconds, with a median (resp. average) running time of $20.0$ (resp. $23.5$) seconds. 
\todo{The other two programs were not able to complete the transformation task within one hour due to inefficiencies in the generated code.}
%Our further investigation shows that these two programs contain nested loops and complex predicates which significantly degrade their efficiency.}
%\todo{We set the 5 minutes time out to ensure that the synthesized program can perform the desired transformation task efficiently. The $2$ programs which were not able to complete their transformation task within the time limit contain nested loops and complex predicates which significantly degrades their efficiency.}

%In order to evaluate the performance of the synthesized programs by \mitra, we applied the returned program for each XML benchmark to a sample XML document of size $512 \pm 20$ MB generated using~\cite{oxygen_xml}. Among all the $48$ solved benchmarks, only $2$ of them were not able to complete the migration task within a $300$ seconds time limit. All the other $46$ programs transformed the XML document to the desired relational table within $8.55$ to $65.51$ seconds, with the median and average times of $19.95$ and $23.47$ seconds.} 

\subsection{Migration to Relational Database}\label{sec:eval-full-db}

\todo{
{\bf \emph{Setup.}}
}
In our next experiment, we use \mitra to convert real-world XML and JSON datasets to a complete relational database. Our benchmarks for this experiment include the following four well-known datasets:

\vspace{-0.05in}
\begin{itemize}[noitemsep]
\item {\bf DBLP}, an XML document containing $2$ GB of data about published computer science papers~\cite{dblp}. 
\item {\bf IMDB}, a set of JSON documents containing $6.2$ GB of data about movies, actors, studios etc.~\cite{imdb}~\footnote{The raw data from~\cite{imdb} is provided in tab-separated-values (TSV) format, so we converted it to JSON format using an existing program~\cite{imdb2json}.}
\item {\bf MONDIAL}, an XML document containing $3.6$ MB of geographical data~\cite{mondial}. 
\item {\bf YELP}, a set of JSON documents containing $4.6$ GB of data about businesses and reviews~\cite{yelp}. 
\end{itemize}

We obtained the target relational database schema for each of these datasets from~\cite{database_schemas}\todo{, and certified that all of them are normalized}. To use \mitra to perform the desired data migration task, we manually constructed small input-output examples. \todo{For each table, we provided a single pair of input-output examples, in addition to a list of all primary and foreign keys in the database schema.} In particular, the average number of elements in the input examples is $16.6$ and the average number of rows in each database table is $2.8$.  Given a suitable input-output example for each target database table, we then ran \mitra with a time limit of $120$ seconds. We then manually inspected the synthesized program and verified its correctness. In this experiment, there was no user interaction; the original examples we provided were sufficient for \mitra to synthesize the desired program. Furthermore, for each target table, we were often able to re-use the input examples (e.g., for IMDB, we used 2 different input examples for the 9 target tables).

\vspace{4pt}\noindent
\todo{
{\bf \emph{Results.}}
}
The results  of this experiment are summarized in Table~\ref{tab:result-fullDB}. The columns under ``Database"€ show the number of tables and total number of attributes in the target database. In particular, the number of database tables range between $7$ and $25$ and the number of columns range from $34$ to $120$.~\footnote{\todo{The Mondial database consists of a large number of tables and columns because it includes a variety of different geographical and cultural data.}} Next, the two columns under ``Synthesis" show total and average synthesis time in seconds, respectively. Specifically, the average time denotes synthesis time per table, whereas total time aggregates over all database tables.  As shown in Table~\ref{tab:result-fullDB}, average synthesis time ranges between $0.8$ to $3.7$ seconds per database table.

The last part of Table~\ref{tab:result-fullDB} provides information about the execution time of the synthesized program and size of the generated database. Specifically, according to the ``\#Rows" column, the total number of rows in the generated database ranges between $27,000$ and $51$ million. The next two rows provide statistics about the execution time of the synthesized programs on the  original full dataset. For example, the average time to generate a target database table for datasets in the range $2-6$ GB is between $31$ and $148$ seconds. These statistics show that the programs generated by \mitra can be used to migrate real-world large datasets.

\vspace{-0.12in}

\section{Related Work}\label{sec:related}

This paper is related to a long line of work in databases and programming languages. In what follows, we compare and contrast our approach against existing techniques.

\vspace{4pt}
\noindent
{\bf Data Exchange.}
The problem of converting hierarchically structured documents to a relational format is a form of \emph{data exchange} problem, where the goal is to transform a data instance of a source schema into a data instance of a target schema~\cite{fagin05}. Due to the difficulty of manually performing such transformations, there has been significant work on automating data exchange tasks~\cite{popa2002translating,clio, roth13, miller2000schema, fagin2005data}. A common approach popularized by \emph{Clio}~\cite{popa2002translating} decomposes the data exchange task into two separate phases: \emph{schema mapping} and \emph{program generation}. A schema mapping describes the relationship between the source and target schemas and is typically specified in the form of declarative logic constraints, such as GLAV (Global-and-Local-As-View) constraints~\cite{kolaitis2005schema}. Given a schema mapping, the second program generation phase ``translates" this mapping into executable code~\cite{clio}.

Because manual construction of schema mappings requires non-trivial effort from data architects, there has been significant work on automatically inferring such mappings from various kinds of informal specifications provided by the user. For instance, users may specify element correspondences by drawing lines between elements that contain related data~\cite{clio,kang2003schema, madhavan2001generic, do2002coma, nandi2009hamster, elmeleegy2008usage}. 
More recently, examples have become more popular as a way of communication between users and the system.
Such example-based approaches can be broadly categorized into two classes depending on what the examples are used for. 
One approach uses examples to specify the desired schema mapping~\cite{qian2012sample, alexe2011designing, alexe2011characterizing, alexe2011eirene}.
%as the specification for learning schema mappings, where users give example instances to the system as inputs~\cite{qian2012sample, alexe2011designing, alexe2011characterizing, alexe2011eirene}.
To the best of our knowledge, all of these approaches use GLAV (or similar formalisms) as the schema mapping language, and hence can only handle cases where the source and target schemas are both relational. 
The other approach uses examples to help users understand and refine the generated schema mappings~\cite{yan2001data, alexe2008muse}, and the learning procedure still takes visual correspondences (lines between elements) as specification. 
Our work can be viewed as an instantiation of the first approach. However, our method is different from previous methods of this approach in three important aspects. First, we can synthesize programs that convert tree-structured  to relational data. Second, we design a DSL as the intermediate language which can express mappings between hierarchical and relational formats. Finally, our PBE-based synthesis algorithm combines finite automata and predicate learning to enable efficient mapping generation.

\begin{comment}
Other approaches infer schema mappings based on user-provided examples~\cite{, , , , , }. This line of work leverages the observation  that users usually understand their data better than schema mappings and can therefore successfully demonstrate their intent using data examples~\cite{alexe2008muse}. Our work is based on the same observation; however, our approach directly constructs an executable transformation script instead of inferring an intermediate schema mapping.

\todo{Say something about pros and cons of our approach vs. theirs. Ideally, we'd have some empirical data to back this up.}
\end{comment}

\vspace{4pt}
\noindent
{\bf Programming-by-Example (PBE).}
The problem of automatically synthesizing programs that satisfy a given set of input-output examples has been the subject of research in the past four decades~\cite{shaw1975}. Recent advances in algorithmic and logical reasoning techniques have led to the development of PBE systems in a variety of domains including string transformations~\cite{flashfill,blinkfill}, data filtering~\cite{fidex}, data imputation~\cite{dace}, data structure manipulations~\cite{l2,hades}, matrix transformation~\cite{syngarextended}, table transformations~\cite{morpheus,harris2011spreadsheet}, SQL queries~\cite{scythe,sqlsynthesizer}, and map-reduce distributed programs~\cite{mapreduce}.

%Among existing PBE techniques, one particularly related work is {\sc Hades}~\cite{hades}, which can be used to synthesize transformations between tree-structured data, including XML files, from input-output examples. While {\sc Hades} uses a similar internal representation as our HDTs, {\sc Hades} cannot automate transformations from hierarchical to relational data. 
\todo{
Among existing PBE techniques, one particularly related work is {\sc Hades}~\cite{hades}, which can be used to synthesize transformations between tree-structured data, including XML files, from input-output examples. While our internal HDT representation is inspired by {\sc Hades}, there are some subtle differences: In particular,  instead of mapping each element to a single node as in \textsc{Hades}, we map each element to multiple tree nodes.
Despite the similarity between the HDT representations, \textsc{Hades} focuses on tree-to-tree transformations and  cannot automate transformations from hierarchical to relational data. Although a relational table can be represented as an HDT, {\sc Hades}' approach of decomposing trees to a set of paths omits  relations between different paths (i.e., columns in the relational table) and can therefore not automate most kinds of interesting tree-to-table transformations.
}
To the best of our knowledge, the only prior PBE-based tool that aims to automate transformations from semi-structured to relational data is {\sc FlashExtract}~\cite{flashextract}. While the main focus of {\sc FlashExtract} is data stored in spreadsheets, it also supports some tree-structured data, such as HTML documents. However, {\sc FlashExtract} is less expressive than \mitra, as it cannot infer relations between different nodes in the tree structure.

%Previous work~\cite{hades} has studied the problem of automatically transforming tree-structured data into tree-structured data from input-output examples. In this work, we study a new and harder problem of transforming semi-structured data into relational databases using PBE techniques. Our main novelty is that we combine finite automata and predicate learning in our synthesis algorithm. 

\vspace{4pt}
\noindent
{\bf XML-to-Relational Mapping.} There is a significant body of research on using relational database management systems to store and query XML (and JSON) documents~\cite{krishnamurthy2004xml,shanmugasundaram2001general,dweib2008schemaless,yoshikawa2001xrel,jiang2002xparent, tatarinov2002storing, shanmugasundaram1999relational, ShreX, atay2007efficient}. A typical approach for this problem consists of three steps: First, the tree structure of the document is converted into a flat, relational schema; next, the XML document is ``shredded" and loaded into the relational database tables, and, finally, XML queries are translated into corresponding SQL queries. The goal of these systems is quite different from ours: In particular, they aim to efficiently answer queries about the XML document by leveraging RDBMS systems, whereas our goal is to answer SQL queries on a desired relational representation of the underlying data. 
\todo{
Furthermore, our XML-to-Relational mapping algorithm is also different from existing XML shredding techniques which can be broadly categorized into two classes, namely structure-centric and schema-centric approaches. 
The first approach relies on the XML document structure to guide the mapping process~\cite{tatarinov2002storing, soltan2006clustering}, whereas the second one makes use of schema information and annotations to derive a mapping~\cite{ShreX, annot1, annot2}. 
In constrast, our technique uses instances of the input XML document and the output relational table in the mapping process. 
Moreover, we have designed a novel DSL that can express a rich class of mapping programs and is amenable to effient synthesis. 
Finally, the combination of techniques based on DFA and predicate learning in our synthesis algorithm is different from previous XML shredding approaches. 
}

%XML shredding is the task of converting XML documents into relational tables~\cite{krishnamurthy2004xml}. XML shredder is a necessary building block of an XML storage and querying system that uses relational databases as its backend~\cite{shanmugasundaram2001general}. In order to store an XML document in relational databases, the tree structure of the XML document must first be converted into an equivalent, flat, relational schema. XML documents are then shredded and loaded into the relational tables. 
%The XML shredding techniques can generally be categorized into two classes. 
%$The first class is schemaless approach which uses the XML document structure (instead of the schema) in the shredding process~\cite{dweib2008schemaless, yoshikawa2001xrel, jiang2002xparent, tatarinov2002storing}. 
%The second class is the schema-based approach which makes use of schema information~\cite{shanmugasundaram1999relational, ShreX, atay2007efficient}. 
%Our technique can be viewed as an instance of the schemaless approach, where we do not require the end-user to provide any schema of the XML document. Instead, our technique uses the structure as well as data information in the examples to search for the desired transformation. 

\vspace{-0.04in}
\section{Conclusions and Future Work}\label{sec:conclusion}

In this paper, we proposed a new PBE-based method  for  transforming hierarchically structured data, such as XML and JSON documents, to a relational database. Given a small input-output example, our tool, \mitra, can synthesize XSLT and Javascript programs that perform the desired transformation. The key idea underlying our method is to decompose the synthesis task into two phases for learning the column and row extraction logic separately. 
%Our synthesis algorithm generates column extractors by constructing a certain type of DFA and identifying the lowest-cost word accepted by this DFA. In contrast, our method learns the predicates used in the row extraction logic by reducing the problem to a combination of integer linear programming and logic minimization. 
We have evaluated our method on examples collected from Stackoverflow as well as real-world XML and JSON datasets. Our evaluation shows that \mitra is able to synthesize the desired programs for $93\%$ of the Stackoverflow benchmarks and for all of the real-world datasets.  

For future work, we are interested in further optimizing the data migration programs generated by \mitra. In particular, our method performs data migration by running a separate program per database table. However, because these programs operate over the same dataset, we can reduce overall execution time by memoizing results across different programs. 
%\todo{Also, we would like to improve the system to overcome its limitations which are discussed in Section~\ref{sec:eval-accuracy}. For instance, when the desired program for a set of examples is not expressible in our DSL, \mitra can automatically divide the examples to several smaller pieces such for each of those it can find a transformation program in our DSL.}
\todo{
Another limitation is that our synthesis algorithm does not return anything if there is no DSL program that satisfies the given examples. To provide more meaningful results to the user in such cases, we  plan to investigate techniques that can synthesize programs that maximize the number of satisfied input-output examples.}

\newpage
\balance

\bibliographystyle{abbrv}
\bibliography{main}

\newpage
%\newpage

%\section*{Appendix }\label{app:proofs}
\appendix

\section{Proofs of Theorems}\label{appendix:proofs}

\vspace{0.1in}\noindent
{\bf \emph{Proof of Theorem~\ref{thm:fta}.}}

We first prove the soundness of $\fta$ -- i.e., if $\fta$ accepts a column extraction program $\cextract$ in our DSL, then we have $\forall (\extree, \rel) \in \ex. \ \lbb \cextract \rbb_{\{ \extree.\text{root} \}, \extree} \supseteq \emph{column}(\rel, i)$. Recall that $\fta$ is the intersection of each $\fta_j (j = 1, \dots, |\ex|)$ constructed using rules shown in Figure~\ref{fig:fta-rules}. We prove $\fta$ is sound by proving the soundness of each $\fta_j$. Since rule (5) in Figure~\ref{fig:fta-rules} marks a state $q_s$ to be a final state only if we have $s \supseteq column(\mathcal{R}, i)$ and any program $\pi$ accepted by $\fta_j$ must evaluate to the value $s$ in a final state $q_s$, we have that any program $\pi$ accepted by $\fta_j$ satisfies example $\ex_j$. Therefore, we have proved the soundness of $\fta$ also holds.

Now, we prove the completness of $\fta$ -- i.e., if there exists a DSL program $\pi$ such that we have $\forall (\extree, \rel) \in \ex. \ \lbb \cextract \rbb_{\{ \extree.\text{root} \}, \extree} \supseteq \emph{column}(\rel, i)$, then $\pi$ is accepted by $\fta$. We also prove this by showing the completeness of each $\fta_j (j = 1, \dots, |\ex|)$ which is constructed using rules in Figure~\ref{fig:fta-rules}. This can be proved because (a) the construction rules exhaustively apply all the DSL operators until no more value can be produced by any DSL program, and (b) any state $q_s$ such that we have $s \supseteq column(\mathcal{R}, i)$ is marked as a final state. Therefore, we have proved the completeness of $\fta$.

\vspace{0.2in}\noindent
{\bf \emph{Proof of Theorem~\ref{thm:predicate}.}}

Given a set of examples $\ex$ and a table extractor $\psi$, assume there exists a formula $\varphi$ (a boolean combination of atomic predicates) in our predicate language such that we have $\forall (\extree, \rel) \in \ex. \ \lbb \emph{filter}(\textract, \lambda t. \varphi) \rbb_{\extree} = \rel$, now we prove that Algorithm 3 returns a formula $\phi$ such that (a) we have $\forall (\extree, \rel) \in \ex. \ \lbb \emph{filter}(\textract, \lambda t. \pred) \rbb_{\extree} = \rel$, and (b) formula $\phi$ has the smallest number of unique atomic predicates.

First, given a set $\preds^*$ of atomic predicates and suppose there exists a formula constructed as a boolean combination of atomic predicates in $\preds^*$ that can filter out all spurious tuples, it is obvious that Algorithm 3 (lines 12--14) is guaranteed to find such a formula $\phi$.

Then, we show that the call to the {\sc FindMinCover} procedure (given in Algorithm 4) at line 11 in Algorithm 3 returns a set $\preds^*$ of atomic predicates such that (1) there exists a boolean combination of predicates in $\preds^*$ that can filter out all spurious tuples, and (2) the set $\preds^*$ is smallest. In particular, condition (1) holds because Algorithm 4 learns all the necessary atomic predicates for differentiating the positive examples and negative examples among all possible atomic predicates in our DSL, and conditional (2) holds due to our ILP formulation (lines 7--10).

Now, we have proved that Algorithm 3 returns a formula $\phi$ that is a valid filtering formula and has the smallest number of atomic predicates.

\vspace{0.2in}\noindent
{\bf \emph{Proof of Theorem~\ref{thm:soundness}.}}
Given a set of input-output examples $\ex$, we show that for each example $\ex_j$: (a) the learnt column extraction program for column $i$ overapproximates column $i$ in the output table, (b) the learnt table extraction program overapproximates the output table, and (c) the synthesized filter program returns exactly the output table.

First, condition (a) holds due to \textsc{Theorem 1}. Condition (b) also holds because our synthesis algorithm constructs the table extraction program by taking the cross-product of all column extraction programs. Finally, condition (c) holds because of \textsc{Theorem 2}. Therefore, we conclude the proof.

%Theorem~\ref{thm:fta} guarantees the soundness of column extraction programs generated in line $5$ of Algorithm~\ref{alg:synth}:\\
 %\sloppy$ \forall \cextract_j \in \cextracts_j. \ \forall (\extree, \rel) \in \ex.\ \lbb \cextract_j \rbb_{\{ \extree.\text{root} \}, \extree} \supseteq \emph{column}(\rel, i)$. \\
 %Since $\textracts$ is generated as the cross-product of column extraction programs $\cextracts_j$ for every column in the table, any table extraction program $\textract \in \textracts$ satisfies the requirement \sloppy$\forall (\extree, \rel) \in \ex. \ \rel \subseteq \lbb \textract \rbb_{\extree}$ of Theorem~\ref{thm:predicate}. Therefore, Theorem~\ref{thm:predicate} ensures that any predicate generated in line $9$ of the algorithm satisfies the property \sloppy$\forall (\extree_i, \rel_i) \in \ex. \ \lbb \emph{filter}(\textract, \lambda t. \pred) \rbb_{\extree_i} = \rel_i$. Hence, \sloppy$\forall  (\extree_i, \rel_i) \in \ex. \ \lbb P \rbb_{\extree_i} = \rel_i$ is correct for any program $P$ generated and returned in algorithm~\ref{alg:synth}.

\vspace{0.2in}\noindent
{\bf \emph{Proof of Theorem~\ref{thm:completeness}.}}
This theorem holds because (1) the completeness of column extraction program synthesis (as proved in \textsc{Theorem 1}), and (2) the completness of the predicate learning algorithm (as proved in \textsc{Theorem 2}).

%If a program $P$ in our DSL exists such that \sloppy$\forall i \in [1, m]: \ \lbb P \rbb_{\extree_i} = \rel_i$, this means there exist a table extraction program $\textract^*$ and a predicate $\pred^*$ (both expressible in our DSL) such that $P := \lambda \tree. \emph{filter}(\textract^*, \lambda t. \pred^*)$.
%Since theorem~\ref{thm:fta} guarantees that the {\sc LearnColExtractors} procedure generates all the possible column extraction programs $\cextracts_j$ for every column $j$, and their cartesian product returns all the table extractor programs $\textracts$ expressible in our DSL, therefore $\textract^* \in \textracts$.
%The loop in lines $8-12$ of Algorithm~\ref{alg:synth} calls the procedure {\sc LearnPredicate} for every $\textract \in \textracts$, including $\textract^*$. Since $\textract^*$ satisfies the requirement of theorem~\ref{thm:predicate} (see the proof of Theorem~\ref{thm:soundness}) and the formula $\pred^*$ exists in our DSL, Theorem~\ref{thm:predicate} ensures that the predicate learning algorithm returns a predicate $\pred$ for $\textract^*$ such that \sloppy$\forall (\extree, \rel) \in \ex. \ \lbb \emph{filter}(\textract^*, \lambda t. \pred) \rbb_{\extree} = \rel$. Hence, the {\sc LearnTransformation} procedure will return a program $P^* \neq \bot$ after the termination of the loop.

\vspace{0.2in}\noindent
{\bf \emph{Proof of Theorem~\ref{thm:simplicity}.}}
Our synthesis algorithm (shown in Algorithm 1) iterates over all possible table extraction programs (line 8). For each candidate table extraction program $\psi$, our algorithm learns the smallest formula $\phi$ (line 9) and constructs a corresponding filter program (line 11). Because our cost function $\theta$ always assigns a lower cost to a smaller formula, the learnt filter program in each iteration is guaranteed to have the lowest cost among all the filter programs that use the same table extraction program. Finally, the synthesized filter program $P^*$ (line 13) has the lowest cost due to the update at line 12. Therefore, we conclude the proof.

%As we mentioned in the proof of Theorem~\ref{thm:completeness}, line $6$ of Algorithm~\ref{alg:synth} generates all the table extractor programs $\textracts$ expressible in our DSL. For each table extractor $\textract \in \textracts$, the {\sc LearnPredicate} procedure returns a formula $\pred$ with the minimum number of atomic predicates (see the proof of Theorem~\ref{thm:predicate}). Therefore, line $11$ constructs the minimum cost program(w.r.t $\theta$) for a given table extractor $\textract$. Finally, the if statement in line $12$ iterates over all the generated programs and returns a program $P^*$ with the minimum cost. 

%Note that the cost function $\theta$ assigns scores to programs $P_1$ and $P_2$ based on the number atomic predicates they use. Therefore, the guarantee that the {\sc LearnPredicate} procedure returns a minimum cost formula (w.r.t. $\theta$) for a given table extractor $\textract$ comes from the fact that the {\sc FindMinCover} procedure returns the minimum set of predicates for $\textract$. 

\vspace{0.25in}
\section{Empirical Complexity}\label{appendix:complexity}
Let $m$ be the number of input-output examples, $k$ and $r$ be (respectively) the number of columns and maximum number of rows in output tables, and $n$ be the maximum number of nodes in an input tree. Despite the worst-case exponential time complexity in theory, the empirical complexity of our algorithm is close to $O(m^2 \cdot k^2 \cdot n^2 \cdot r \cdot (\log n)^k)$. To see where this result comes from, we first discuss the complexity of the {\sc LearnColExtractor} procedure. The complexity of DFA construction is $O(n)$, and the complexity of generating the intersection of two DFAs is $O(n^2)$. Therefore, the overall complexity of Algorithm~\ref{alg:extractor} is $O(m \cdot n^2)$, and as the result, the loop in lines $4-5$ of Algorithm~\ref{alg:synth} takes $O(k \cdot m \cdot n^2)$. Our empirical evaluations show that the cartesian-product generated in line $6$ of the  {\sc LearnTransformation} procedure usually contains a small number of table extractors. 

Now we discuss the complexity of the {\sc LearnPredicate} procedure for a given table extractor $\textract$ and the set of examples $\ex$. It first generates the universe of all atomic predicates which contains $O(k^2 \cdot n^2)$ predicates. Then, it constructs the set of positive and negative examples. The number of positive examples is determined by the given input-output examples, which is $O(m \cdot r)$. The number of negative examples for each $(\extree, \rel) \in \ex$ is bounded by the number of tuple in the corresponding intermediate table ($ \lbb \textract \rbb_{\extree}$). In practice, each column extractor usually extracts $O(\log n)$ nodes from a given tree with $n$ nodes, therefore the intermediate table contains $O((\log n)^k)$ tuples. Hence, the total size of negative examples is $O(m \cdot (\log n)^k)$. In our experiments, Algorithm~\ref{alg:cover} returned a smallest set of atomic predicates, which included a few predicates, in $O(m^2 \cdot k^2 \cdot n^2 \cdot r \cdot (\log n)^k)$. Finally, finding a smallest classifier over the predicates $\preds^*$ returned by the {\sc FindMinCover} procedure is very efficient because of the small size of $\preds^*$. Therefore, the overall complexity of Algorithm~\ref{alg:pred} is close to $O(m^2 \cdot k^2 \cdot n^2 \cdot r \cdot (\log n)^k)$ in practice. Going back to Algorithm~\ref{alg:synth}, since the loop in lines $8-12$ iterates over a small number of column extractors, the overall empirical complexity of our synthesis algorithm is determined by the complexity of the Algorithm~\ref{alg:pred}.

%However, \emph{in practice}, the complexity of our algorithm is close to $O(m^2 \cdot k^2 \cdot n^2 \cdot r \cdot (\log n)^k)$, where $m$ is the number of input-output examples, $k$ and $r$ are (respectively) the number of columns and maximum number of rows in output tables, and $n$ is the maximum number of nodes in an input tree. More precisely, the {\sc LearnColExtractor} function takes $O(m \cdot n^2)$ time to generate column extractors for each column, and a small number of table extractors are generated in line $8$ of Algorithm~\ref{alg:synth}. The bottleneck of our system is the {\sc LearnPredicate} procedure, which generates $O(k^2 \cdot n^2)$ atomic predicates, $O(m \cdot r)$ positive examples, and $O(m \cdot (\log n)^k)$ negative examples. This results in an empirical complexity of $O(m^2 \cdot k^2 \cdot n^2 \cdot r \cdot (\log n)^k)$ for Algorithm~\ref{alg:pred}. % which determines the overall complexity of our synthesis algorithm.

\vspace{0.25in}
\section{Program Optimization}\label{appendix:optimization}
Our DSL decomposes a tree-to-table transformation task into two subproblems, namely (i) generating a set of column extractors and (ii) learning a filtering predicate. While this DSL is chosen intentionally to facilitate the synthesis task, the resulting programs may be inefficient. In particular, programs represented in this DSL first extract all possible values for each column of the table from the input tree, then generate all possible tuples as the cross-product of those columns, and finally remove all spurious tuples. 
The goal of the program optimization step is to apply the predicate as early as possible to avoid the generation of undesired tuples rather than filtering them out later.
%In this way, the optimized program directly generates the desired table without any spurious tuples. 

The program optimization step in \mitra works as follows: Consider a synthesized program of the form $ \lambda \tree. \emph{filter}(\textract, \lambda t. \pred)$ where $\textract = \cextract_1 \times \ldots \times \cextract_k$. Without any optimization, this program would be implemented as:
\[
\begin{array}{l}
R \ = \ \emptyset \\
for \ n_1 \in \cextract_1(T) \\
\ \ \ \ \ \ \ for \ n_2 \in \cextract_2(T) \\
\ \ \ \ \ \ \ \ \ \ \  \ldots \\
\ \ \ \ \ \ \ \ \ \ \ \ \ \ \ \ for \ n_k \in \cextract_k(T) \\
\ \ \ \ \ \ \ \ \ \ \ \ \ \ \ \ \ \ \ \ \ if(\pred((n_1, n_2, \ldots, n_k)))\\
\ \ \ \ \ \ \ \ \ \ \ \ \ \ \ \ \ \ \ \ \ \ \ \ \ R \ = \ R \cup \{(n_1, n_2, \ldots, n_k)\}
\end{array}
\]

The optimizer first converts the predicate $\pred$ to a CNF formula $\pred = \pred_1 \land \ldots \land \pred_m$. It then generates two formulas $\psi$ and $\chi$ such that $\psi$ is used to guide the optimization, and $\chi$ corresponds to the remaining filtering predicate that is not handled by the optimization. Specifically, $\psi, \chi$ are both initialized to true and updated as follows:: For each clause $\pred_i$ of the CNF formula, we conjoin it with $\psi$ if it is of the form $( (\lambda n. \nextract_1) \ \tuple[i] ) \logicop ( (\lambda n. \nextract_2) \ \tuple[j] )$  and with $\chi$ otherwise.
 
Now, we use formula $\psi$ to optimize the program by finding shared paths of different column extractors. For each $\pred_k = ( (\lambda n. \nextract_1) \ \tuple[i] ) \logicop ( (\lambda n. \nextract_2) \ \tuple[j]  \in \psi$ and extractors $\pi_i, \pi_j$ for columns $i$ and $j$, the optimizer generates two extractors $\cextract^{'}_{i} = \nextract_1(\cextract_i)$ and $\cextract^{'}_{j} = \nextract_2 (\cextract_j)$.  Then, it checks whether $\cextract^{'}_{i}$ and $\cextract^{'}_{j} $ are semantically equivalent programs. If they are not equivalent, the optimizer removes $\pred_k$ from $\psi$ and  conjoins it with $\chi$. If $\cextract^{'}_{i}$ is equivalent to $\cextract^{'}_{j}$ and it is a prefix of both  $\cextract_i$ and $\cextract_j$, then we can represent $\cextract_i = \cextract^{i}_{\emph{suffix}}(\cextract^{'}_{i})$ and $\cextract_j = \cextract^{j}_{\emph{suffix}}(\cextract^{'}_{i})$. This transformation allows us to share the execution of $\cextract^{'}_{i}$ for columns $i$ and $j$ and has the following two benefits: First, it eliminates redundant computation, and, even more importantly,  it guarantees that extracted nodes $n_i$ and $n_j$ satisfy predicate $\psi$.  This kind of reasoning allows us to generate the following optimized program:

\[
\begin{array}{l}
R \ = \ \emptyset \\
for \ n_1 \in \cextract_1(T) \\
\ \ \ \  \ldots \\
\ \ \ \ \ \ \ for \ n_{ij} \in \cextract^{'}_{i}(T) \\
\ \ \ \ \ \ \ \ \ \ n_i = \cextract^{i}_{\emph{suffix}}(n_{ij}) \\ 
\ \ \ \ \ \ \ \ \ \ n_j = \cextract^{j}_{\emph{suffix}}(n_{ij}) \\ 
\ \ \ \ \ \ \ \ \ \ \ \ \  \ldots \\
\ \ \ \ \ \ \ \ \ \ \ \ \ \ \ \ for \ n_k \in \cextract_k(T) \\
\ \ \ \ \ \ \ \ \ \ \ \ \ \ \ \ \ \ \ \ \ if(\chi((n_1, n_2, \ldots, n_k)))\\
\ \ \ \ \ \ \ \ \ \ \ \ \ \ \ \ \ \ \ \ \ \ \ \ \ R \ = \ R \cup \{(n_1, n_2, \ldots, n_k)\}
\end{array}
\]

Observe that the optimized code has a single loop for nodes $n_i$ and $n_j$ rather than a nested loop.

\end{document}